\documentclass[12pt,dvipsnames]{iopart}

 \expandafter\let\csname equation*\endcsname\relax 
 \expandafter\let\csname endequation*\endcsname\relax 
\usepackage{amsmath,amsfonts,amssymb}

\usepackage{graphicx,subfloat}
\usepackage{color}
\usepackage{comment}
\usepackage{appendix}
\usepackage{multirow}

\usepackage{epstopdf}
\usepackage{etoolbox}
\usepackage{hyperref}

\makeatletter
\renewcommand\@appendixstar{\@@par
 \ifnumbysec 
 \@addtoreset{table}{section}
 \@addtoreset{figure}{section}\fi
 \setcounter{section}{0}
 \setcounter{subsection}{0}
 \setcounter{subsubsection}{0}
 \setcounter{equation}{0}
 \setcounter{figure}{0}
 \setcounter{table}{0}
 \def\thesection{\Alph{section}} % this line has been \def\thesection{Appendix \Alph{section}} before
 \def\theequation{\ifnumbysec
      \Alph{section}.\arabic{equation}\else
      \Alph{section}\arabic{equation}\fi}
 \def\thetable{\ifnumbysec
      \Alph{section}\arabic{table}\else
      A\arabic{table}\fi}
 \def\thefigure{\ifnumbysec
      \Alph{section}\arabic{figure}\else
      A\arabic{figure}\fi}}
\makeatother
\begin{document}

\graphicspath{{figs/}}
%============================= 
%RACCOURCIS ET META COMMANDES
%================================

\newcommand{\pg}{\paragraph{}}
\newcommand{\spg}{\subparagraph{}}
\newcommand{\dsp}{\displaystyle}

% Equations
\newcommand{\be}{\begin{equation}}
\newcommand{\ee}{\end{equation}}
\newcommand{\ben}{\begin{equation*}}
\newcommand{\een}{\end{equation*}}
\newcommand{\ba}{\begin{align}}
\newcommand{\ea}{\end{align}}
\newcommand{\ban}{\begin{eqnarray*}}
\newcommand{\ean}{\end{eqnarray*}}
\newcommand{\tieq}[1]{\quad \text{#1} \quad} 		
\newcommand{\sieq}{\quad}

% Lettres Speciales
%Calligraphiees
\newcommand{\mA}{\mathcal A}
\newcommand{\mC}{\mathcal C}
\newcommand{\mD}{\mathcal D}
\newcommand{\mE}{\mathcal E}
\newcommand{\mF}{\mathcal F}
\newcommand{\mG}{\mathcal G}
\newcommand{\mH}{\mathcal H}
\newcommand{\mJ}{\mathcal J}
\newcommand{\mL}{\mathcal L}
\newcommand{\mM}{\mathcal M}
\newcommand{\mN}{\mathcal N}
\newcommand{\mO}{\mathcal O}
\newcommand{\mP}{\mathcal P}
\newcommand{\mQ}{\mathcal Q}
\newcommand{\mS}{\mathcal S}
\newcommand{\mX}{\mathcal X}
\newcommand{\mW}{\mathcal W}
\newcommand{\mZ}{\mathcal Z}

%caractères gras
\newcommand{\bA}{{\bf A}}
\newcommand{\bB}{{\bf B}}
\newcommand{\bS}{{\bf S}}
\newcommand{\bF}{{\bf F}}
\newcommand{\bj}{{\bf j}}
\newcommand{\bk}{{\bf k}}
\newcommand{\bl}{{\bf l}}
\newcommand{\bn}{{\bf n}}
\newcommand{\bp}{{\bf p}}
\newcommand{\bq}{{\bf q}} 
\newcommand{\bv}{{\bf v}}
\newcommand{\bx}{{\bf x}}
\newcommand{\bw}{{\bf w}}
\newcommand{\bz}{{\bf z}}
\newcommand{\bzo}{{\bf z_0}}

\newcommand{\bom}{{\boldsymbol\omega }}

% Symboles dedies
\newcommand{\al}{ et al }
\newcommand{\cg}{\text{c.g.}}
\newcommand{\phd}{PhD }

% Symboles math
\renewcommand{\ln}{\log}
\newcommand{\mbR}{{\mathbb R}}
\newcommand{\mbN}{{\mathbb N}}
\newcommand{\mbZ}{{\mathbb Z}}
\newcommand{\isp}[1]{[\![#1]\!]}
\newcommand{\pprime}{{\prime \prime}}

\newcommand{\one}[1]{ {\mathbf 1}_{#1}} % fonction indicatrice
\newcommand{\pdr}[2]{ \dfrac{\partial{#1}}{\partial{#2}}}  % partial derivative
\newcommand{\fdr}[2]{ \dfrac{\delta{#1}}{\delta{#2}}}  % fun derivative
\newcommand{\sfdr}[2]{ \frac{\delta{#1}}{\delta{#2}}}  % fun derivative
\newcommand{\ex}{{\bf \hat x}}\newcommand{\ey}{{\bf \hat y}}\newcommand{\ez}{{\bf \hat z}}
\newcommand{\er}{{\bf \hat r}}\newcommand{\etheta}{{\bf \hat {\boldsymbol\theta}}}

\newcommand{\vD}{|\mD|}
\newcommand{\az}{{A}}
\newcommand{\pb}[1]{ \left(#1\right)} % scalar prod

% integration
\newcommand{\dS}{\text{\bf {dS}}}
\newcommand{\dbq}{{\d \bf q}\,} 
\newcommand{\dbp}{{\d \bf p}\,}
\newcommand{\dbr}{{\d \bf r}\,}
\newcommand{\dx}{dx}
\renewcommand{\d}{\text{d}}
\newcommand{\dz}{\text{dz}}
\newcommand{\dt}{\text{dt}}
\newcommand{\x}{\mathbf x}

%Géométrie axisymétrique
\newcommand{\Ri}{{R_{in}}}
\newcommand{\Ro}{{R_{out}}}

%Liens
\newcommand{\rsec}[1]{section \ref{#1}}
\newcommand{\m}[1]{\langle{#1} \rangle}

%Abbrevations
\newcommand{\tdt }{two-dimensional hydro turbulence }
\newcommand{\bmhdt}{two-dimensional magneto hydro turbulence }
\newcommand{\quasi}{quasi two-dimensional }
\newcommand{\axi}{three-dimensional axisymmetric dynamics }
\newcommand{\bmhd}{two-dimensional magnetohydrodynamics }
\newcommand{\bbou}{two-dimensional Boussinesq flows }
\newcommand{\NS}{Navier-Stokes }
\newcommand{\cano}{\text{cano}}
\newcommand{\micro}{\text{micro}}
\newcommand{\kin}{\text{kin}}
\newcommand{\mg}{\text{mag}}
\newcommand{\pol}{\text{pol}}
\newcommand{\tor}{\text{tor}}
\newcommand{\vk}{ Von K\'{a}rm\'{a}n }
\newcommand{\peristrophy}{peristrophy }
\newcommand{\mix}{\text{mix} }
\newcommand{\Ak}{\left\lbrace A_k\right\rbrace}
\newcommand{\RSM}{Robert-Miller-Sommeria }
\newcommand{\ssigma}{\text{sign} (\sigma)}
\newcommand{\sxi}{\text{sign} (\xi)}
\newcommand{\Dpiv}{{{\mathcal D}_\text{piv}}}
% ENVIRONNEMENTS
%\newcommand{\simon}[1]{\footnote{\textcolor{blue}{\underline{Simon Says : }  #1}}}
%\newtheorem[S]{thm}{Theorem}[section]
%\newtheorem[cut=false]{thm}{Theorem}
%\newtheorem[cut=false]{defi}{Definition}
%\newtheorem[cut=false]{corol}{Corollary}
%\newtheorem[cut=false]{algo}{Algorithm}[chapter]

\newcommand{\se}{{\star,\epsilon}} 

\newcommand{\sg}[2]{{\subfloat[#1]{\includegraphics[width=\textwidth]{#2}}}}
\newcommand{\trg}[1]{\includegraphics[width=\textwidth,trim=8cm 5cm 8cm 5cm, clip]{#1}}
%Mots a hyphener
\hyphenation{Me-cha-nics}
\hyphenation{a-xi-sym-me-tric}
\hyphenation{a-xi-sy-mé-trie}
\hyphenation{a-xi-sy-mé-trie}
\hyphenation{ther-mo-ha-li-ne}

\renewcommand{\sg}[3]{\subfloat[#1]{\includegraphics[width=#3\textwidth]{#2}}}
\renewcommand{\trg}[1]{\includegraphics[width=0.24\textwidth,trim=8cm 5cm 8cm 5cm, clip]{#1}}

\renewcommand{\br}{{\bf r}}
%\captionsetup[subfigure]{labelformat=simple}
%\renewcommand{\thesubfigure}{\relax}
%\renewcommand{\thesubfigure}{(\alph{subfigure})}
\newcommand{\avc}[1]{\langle #1\rangle_\circ}
\newcommand{\av}[1]{\langle #1\rangle}

\newcommand{\mDp}{\mD_{\text{piv}}}
\newcommand{\modif}[1]{#1}
\newcommand{\simon}[1]{\textcolor{blue}{\underline{Simon says :} #1}}
\newcommand{\berengere}[1]{\textcolor{red}{\underline{Berengere says :} #1}}
\newcommand{\del}[1]{\textcolor{green}{#1}}

%\title[Ferro-Turbulence]{Coherent cells in turbulent von K\'arm\'an flows : a statistical mechanics perspective (Ferro-turbulence)}

\title[Ferro-Turbulence]{A statistical mechanics framework for the large scales of turbulent von K\'arm\'an flows}

\author[S.Thalabard, B.Saint-Michel, Eric Herbert,F.Daviaud, B.Dubrulle]{Simon Thalabard\dag, Brice Saint-Michel\ddag, Eric Herbert$ \triangle$, Fran\c cois Daviaud\S\  and B\'ereng\`ere Dubrulle\S} 
%\address{\dag Laboratoire Lagrange UMR 7293, Universit\'e de
%  Nice-Sophia Antipolis,\\ CNRS, Observatoire de la C\^ote d'Azur,
%  Bd.\ de l'Observatoire, 06300 Nice, France.}
\address{\dag Department of Mathematics and Statistics, University of Massachusetts, Amherst, MA 01003, USA.}
 \date{\today}
\address{
\ddag Institut de Recherche sur les Ph\'enom\`enes Hors Equilibre, Technop\^ole de Ch\^ateau-Gombert, 49 rue Fr\'ed\'eric Joliot Curie B.P. 146 13 384 Marseille, France.}
\address{
$\triangle$ Universite Paris Diderot - LIED - UMR 8236
Laboratoire Interdisciplinaire des Energies de Demain - Paris, France\\}

\address{\S
Laboratoire SPHYNX, Service de Physique de l'\'Etat Condens\'e, DSM/IRAMIS/SPEC, CNRS UMR 3680, CEA Saclay, 91191 Gif-sur-Yvette, France\\
}

%\ead{berengere.dubrulle@cea.fr}
\ead{simon.thalabard@ens-lyon.org}

\begin{abstract}
In the present paper, recent experimental results on large scale coherent steady states observed in   experimental von K\'arm\'an flows are revisited from a statistical mechanics perspective. The latter  is rooted on two levels of description. 
We first argue that the coherent steady states may be described as the equilibrium states of well-chosen lattice models, that can be used to define global properties of von K\'arm\'an flows, such as their temperatures.  
The equilibrium description is then enlarged,  in order to reinterpret a series of  results about the stability of those steady states, their susceptibility to symmetry breaking,  in the light of a deep analogy with the statistical theory of  Ferromagnetism.
\end{abstract}

\submitto{\NJP}

\maketitle
\tableofcontents
\section{Introduction}
Describing the complexity of turbulent flows with tools from statistical mechanics is  a long-standing dream of theoreticians. In 1949, five years after the publication of its solution for the problem of phase transition in the 2D Ising model, Onsager published a notorious study of the statistical mechanics of the point vortex model \cite{onsager1949statistical}, a special solution of the 2D Euler equations that allows to interpret the emergence of long lived coherent structures in terms of the pairing between vortices mutually interacting through a long range Coulombian potential.  That Onsager chose the special case of 2D turbulence is probably not a coincidence: as soon as 1947, he was aware of the existence of the  dissipative anomaly in 3D flows that precludes the use of classical equilibrium tools such as micro-canonical measures  \cite{eyink2006onsager}. In other words,  the non-vanishment of the energy flux at vanishing viscosity for 3D flows makes 3D turbulence an intrinsically far-from-equilibrium system, which cannot be properly approximated by crudely setting the viscosity to zero in the Navier-Stokes equations.  \modif{In a 2D flow, if one lets the viscosity go to zero while keeping the injection scale fixed, then the dissipative anomaly disappears. This  justifies Onsager's statistical mechanics approach \cite{Duchon2000}}. Since Onsager, the description of 2D turbulence using statistical mechanics has greatly improved. Starting from the seminal work of Kraichnan in the 1960's \cite{kraichnan1967inertial,kraichnan1980two},  Robert, Miller and Sommeria in the 1990's \cite{miller1990statistical,robert1991statistical,robert1992relaxation} and subsequent work from then \cite{michel1994statistical,jordan1997ideal,ellis2004statistical,majda2006nonlinear,bouchet2010invariant,bouchet2014langevin}, the use of statistical mechanics led to a description of the coherent structures that seems to match the observed large scale organization in experimental and  numerical 2D turbulence \cite{Turkington2001,Bouchet2002,Bouchet2012,Qi2013}.\modif{ Yet, the extension of those ideas to 3D flows has until now been proven unfruitful.}

An exception may  come from the special case of von K\'arm\'an (VK) turbulence,  a now classical human size experiment that reaches very large Reynolds numbers of the order of $10^6$ through the stirring of a fluid in between two counter-rotating propellers. At this value, it is generally expected that the turbulence is fully developed with a wide range of interacting scales \cite{frisch1996turbulence}. This was indeed confirmed by previous analysis of the  turbulence properties in the middle shear layer, which evidenced scaling properties and intermittency corrections in agreement with other measurements in fully developed turbulent flows using different geometries \cite{Pinton1994,Arneodo1996,Ravelet2008,Herbert2012}.
Some indications exist though, that the number of effective degrees of freedom \modif{that describe the large scales of } turbulent VK flows is not so large : at Reynolds number around $10^5-10^6$, Poincar\'e maps of the torque exerted by the turbulent flow on each propeller exhibit beautiful attractors and limit cycles \cite{ravelet2005bifurcations,saint2013forcing}. Those features are usually observed in dynamical systems with only three or four degrees of freedom ---\, see \cite{Eckmann1985}.
This suggests that the system could in principle be efficiently described by only a few global quantities and that a statistical mechanics approach could be used to identify  hydrodynamical analogues for ``temperatures'' or ``chemical potentials''. 
%The purpose of the present paper is to provide experimental evidence that the large scale coherent average states observed in VK experiments can indeed be described using the language and  tools from statistical mechanics. 
\modif{The present paper is precisely meant to support this somewhat thought-provoking idea, namely that the large scales of 3D VK turbulent flows can be encompassed in a broad equilibrium statistical mechanics framework.}

 The  starting point of our analysis is the observation that VK turbulence is \emph{not} isotropic. Besides, and as far as the  average flow is concerned, the swirling flows obtained in VK devices seem to provide an example of 3D turbulence with  axial symmetry.   
 As previously discussed in \cite{leprovost2006dynamics,naso2010statistical}, axially symmetric turbulence  is an intermediate case between 2D and 3D turbulence,  for which  equilibrium theories yield non trivial insights \cite{thalabard2013statistical}.
VK turbulence is however not axially symmetric, and the question remains open whether the predictions obtained using an  ``axi-symmetric ansatz'' are relevant to account for the coarse-grained properties of such flows. Preliminary comparisons performed at large Reynolds numbers by \cite{monchaux2006steady,monchaux2007mecanique} suggest that the steady  states of experimental VK flows can be described in terms of  a restricted set of meta-stable equilibria of the 3D axially symmetric Euler equations. 
   The goal of the present paper is to support further this idea and show how the light of the statistical mechanics can be used \modif{both beyond the scope of ideal theories and beyond the scope of strictly 2D flows}, in order to provide a useful framework of analysis.  We will evidence a deep analogy between the  VK steady states and lattice models  of Ferro-magnetism. %We call this analogy ``Ferro-Turbulence''.
To make the analogy vivid,  we stick to the simplest conceptual level compatible with a comprehensive description of the statistical mechanics features observed in VK flows. The reader interested in more technical details will be referred  to the other publications.
The present paper is organized as follows. We first describe the experimental set-up and its symmetries. We briefly recall the properties of the VK steady states, and of their associated bifurcations. We then summarize the outcomes of several statistical theories associated to the ``ideal axially-symmetric fluid''. Those theories are then used beyond their initial scope, in order to develop an analogy between the experimental VK flow and lattice models of ferro-magnetism. Within this analogy, the previously observed VK bifurcations are shown to be reminiscent of second order mean-field transitions, and critical exponents are measured. We conclude by a discussion of our results.

\section{Coarse-grained  description of a VK flow}
\label{sec:coarsegraineddescription}
\subsection{Control Parameters}
\begin{figure}[htb]
\centering
\begin{minipage}{0.79\columnwidth}
  \begin{minipage}{0.55\columnwidth}
 \includegraphics[width=0.99\columnwidth,trim=0cm 0cm 0cm 5cm,clip]{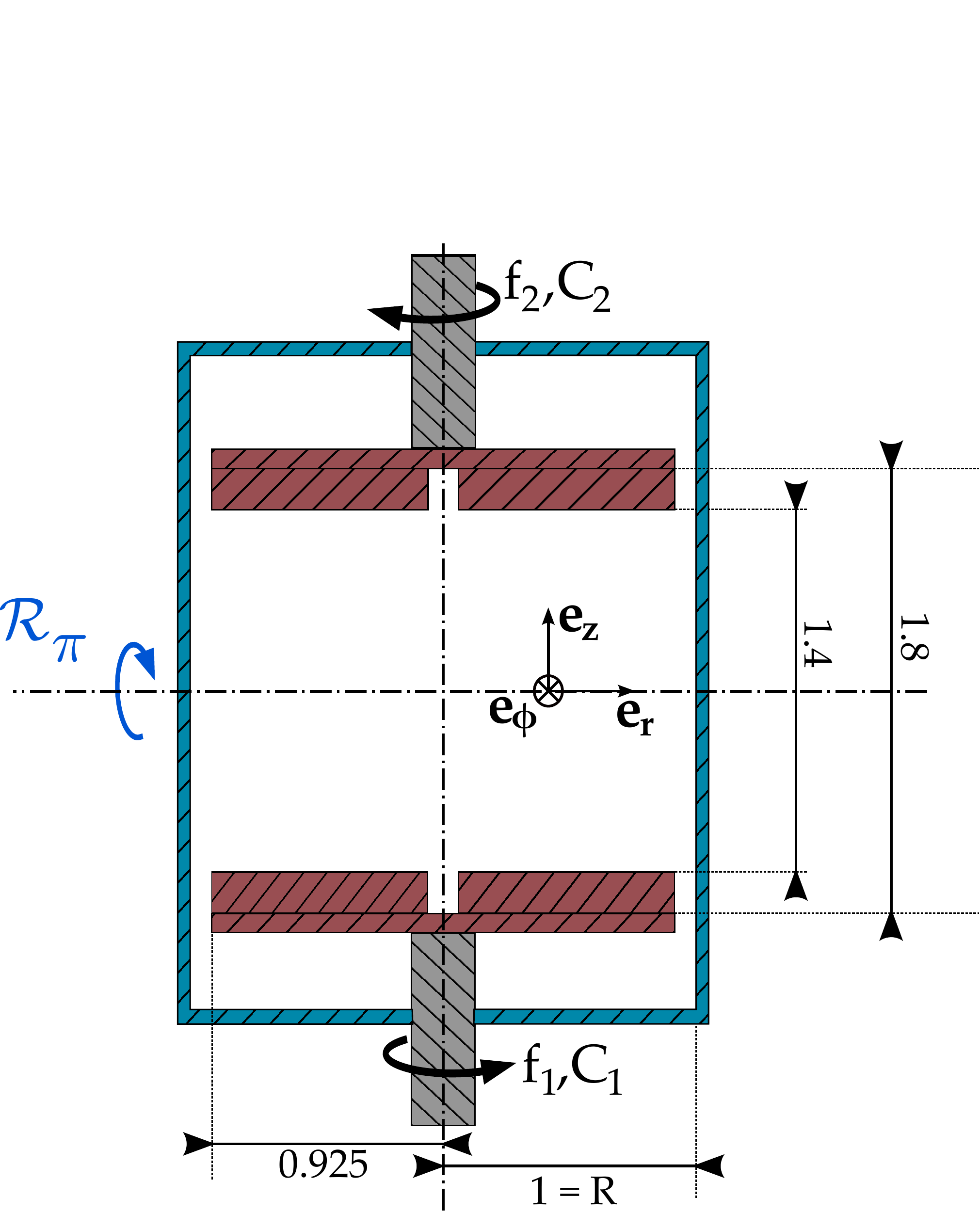}
\end{minipage}
  \begin{minipage}{0.44\columnwidth}
 \includegraphics[width=0.99\columnwidth]{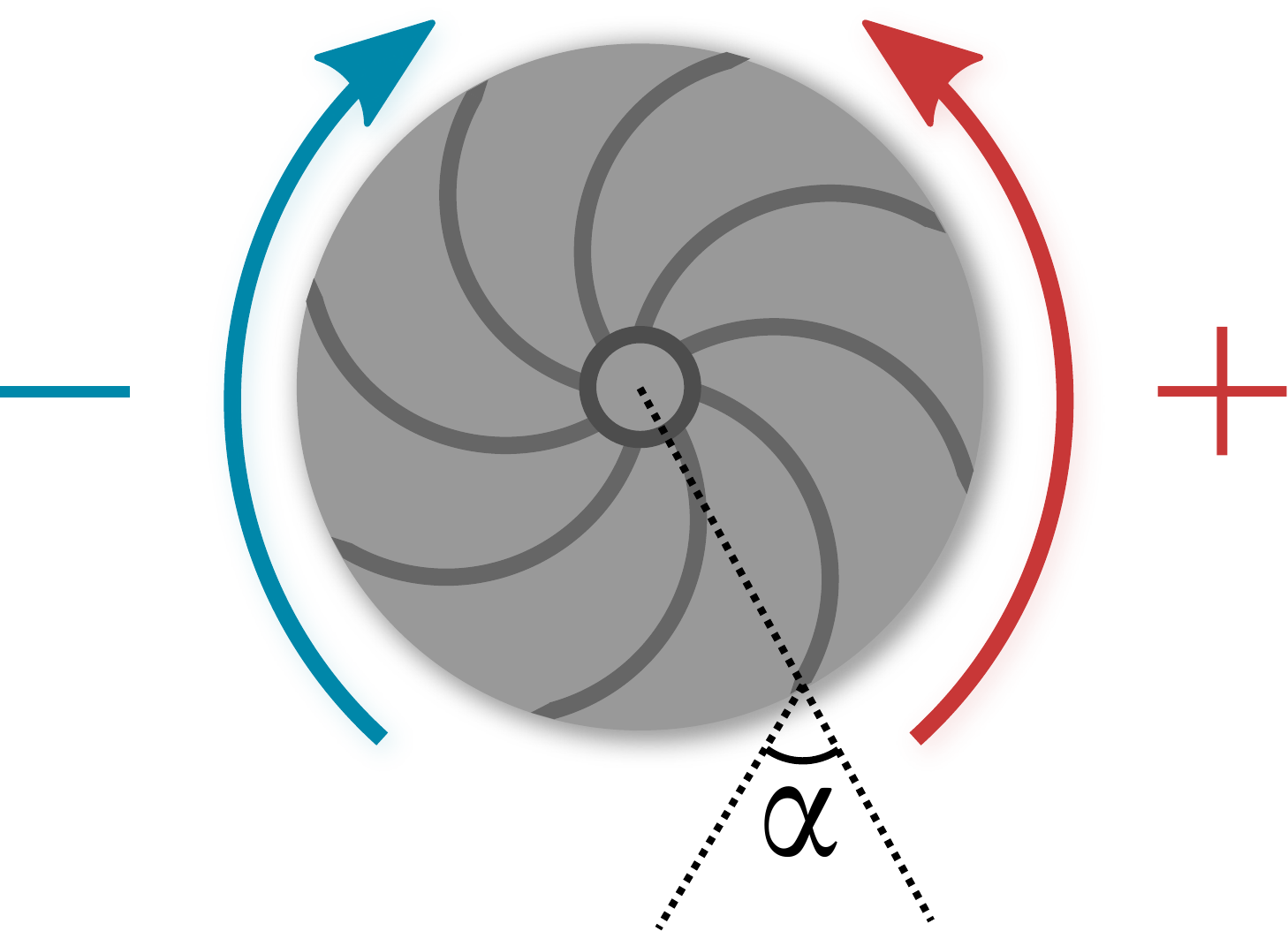}
 \end{minipage}
\end{minipage}
\caption{Left : Sketch of the VK2 experiment. Right: Sketch of a propeller and definition of the oriented angle $\alpha$. }
\label{fig:VI_VK2sketch}
 \end{figure}

The VK experimental set-up used for the present study has been thoroughly described in \cite{ravelet2005bifurcations, monchaux2007mecanique,saint2013these}. The fluid is confined inside a cylinder of radius $R=100$~mm, and forced through two rotating impellers of radius $R_t$ ---\, see  Figure \ref{fig:VI_VK2sketch}. All the lengths will now be expressed in units of the cylinder radius $R$. The aspect ratio of our experiment is  defined as the distance between the inner faces of the two opposite impellers   $2H =1.8$.
Impellers are driven by two independent motors, whose frequencies $f_1$ and $f_2$ can either be set equal, in order to get an exact counter-rotating regime, or set to
different values $f_1\ne f_2$.  To change the viscosity, mixtures of water or glycerol with different dilution rates were used. 

%Figure 1

In this paper,  three  main global parameters are used to  characterize VK turbulence. \emph{(i)} The \emph{Reynolds number} $Re=\pi (f_1+f_2) R^2 \nu^{-1}$ ---\, with  $\nu$  the fluid
kinematic viscosity \,--- ranges from  $10^2$ to $5 \times 10^5$ so that a full range of regimes can be spanned, from a purely laminar to a fully turbulent one. \emph{(ii)} The \emph{rotation number}, $\theta=(f_1-f_2)/(f_1+f_2)$, measures the relative influence of global rotation over a typical turbulent shear frequency.
 It can be varied from $-1$ to $+1$, hereby exploring a regime of relatively weak rotation to shear ratio. \emph{(iii)} Finally, the \emph{torque asymmetry} $\gamma=(C_1-C_2)/(C_1+C_2)$ measures the difference between the torques $C_1$ and $C_2$ applied to each of the propellers. It is crucial to note that in the VK2 experiment, turbulence can be either generated by maintaining constant the frequencies or the torques  applied to each of the propeller ---\, please, see \cite{ravelet2005bifurcations,saint2013forcing} for more details.
 
At a finer level of description, it has been shown that the turbulence properties (anisotropy, fluctuations, dissipation) are influenced by the geometry of the propellers, \emph{viz.},  their non dimensional radius $R_t$, the oriented  angle  $\alpha$  between  the blades and the rotation direction (see  Figure \ref{fig:VI_VK2sketch}), the heights $h_b$  and  the number $n$ of blades \cite{ravelet2005bifurcations}. In the present paper, we consider only propellers with $h_b=0.2$ and focus on changes induced by variations of $\alpha$. Those propellers are the so-called ``TM60'', ``TM87'' and ``TM73'' propellers, whose characteristics are summarized in Table  \ref{table:param}.  
A single  propeller can be used to propel the fluid in two opposite directions, respectively associated to the concave or convex face of the 
blades going forward. This can be accounted by a change of sign of the parameter $\alpha$. In the sequel, we denote $(-)$ ({\it
resp.} $(+)$ ) a propeller used with the concave ({\it resp.} convex) face of its blades going forward.

 Table \ref{table:param} summarizes the parameter space that was explored in our system.  Schematically, the influence of the propeller geometry has been explored at $Re=10^5$, $\theta=0$. The Reynolds variation has been explored at $\theta=0$ using the TM60 propellers ($\pm$).  The rotation variation has been explored at $Re=10^5$ using TM73($\pm$), TM87($\pm$) and TM60($\pm$). The influence of the forcing type (``constant velocity'' against ``constant torque'' forcing) has been studied  with the TM60(-) and TM87(-) at $Re=10^5$.

\begin{table*}[ht]
\centering
\resizebox{0.95\textwidth}{!}{
\begin{tabular}
{||c||c|c|c|c||}%
\hline \hline
    Propellers   &Number of blades&$\alpha$  (in degrees)       &$Re$&$\theta$\\
\hline \hline
All&8 and 16&$ [-90,90]    $&$10^5$&$0 $\\
%\hline
TM60(+)&16&$72$&$[10^2,10^6]$&$ [-1,1]$\\
%\hline
TM60(-)&16&$-72$&$10^5    $&$ [-1,1] $\\
%\hline
TM87(+)&8&$72$&$10^5$&$ [-1,1]$\\
%\hline
TM87(-)&8&$-72$&$10^5    $&$ [-1,1] $\\
%\hline
TM73(+)&8&$+24$&$10^5    $&$ [-1,1] $\\
%\hline
TM73(-)&8&$-24$&$10^5    $&$ [-1,1] $\\
\hline \hline
\end{tabular}
}
{\caption{Parameter space explored in our set-up} \label{table:param}.}
\end{table*}
\subsection{Topology of the averaged steady states}
\subsubsection{A qualititative description.}
To analyze the topology of the averaged states, Stereoscopic Particle Image Velocimetry (SPIV) measurements were mostly used. The system provides the three components of the velocity field on a $95\times66$ points grid, that covers  a whole meridian plane of the flow through time series of about $600$ to $5000$ regularly sampled values at a 10Hz frequency.  We also performed a few Laser Doppler Velocimetry measurements providing mean velocities over a $11 \times 13$ points grid covering a half meridian plane of the flow. 
To deal with non-dimensional velocity fields, those are divided by a typical ``forcing velocity'' defined as  $V_0=2\pi R(f_1+f_2)/2$. We write  $(r,\phi,z)$ the standard cylindrical coordinates, and denote $\langle.\rangle_\circ$ an average over a time series of SPIV measurements.  We also use the short-hand notation  $\int_\Dpiv f = (1/2H)\int_{0}^1 r dr \int_{-H}^H (f(r,0,z)+f(r,\pi,z))$ to denote the spatial average of any quantity $f=f(r,\phi,z)$ over the PIV window.  
\begin{figure}[!htb]
\centering
\begin{minipage}{0.79\textwidth}
\centering
 \includegraphics[width=0.59\columnwidth]{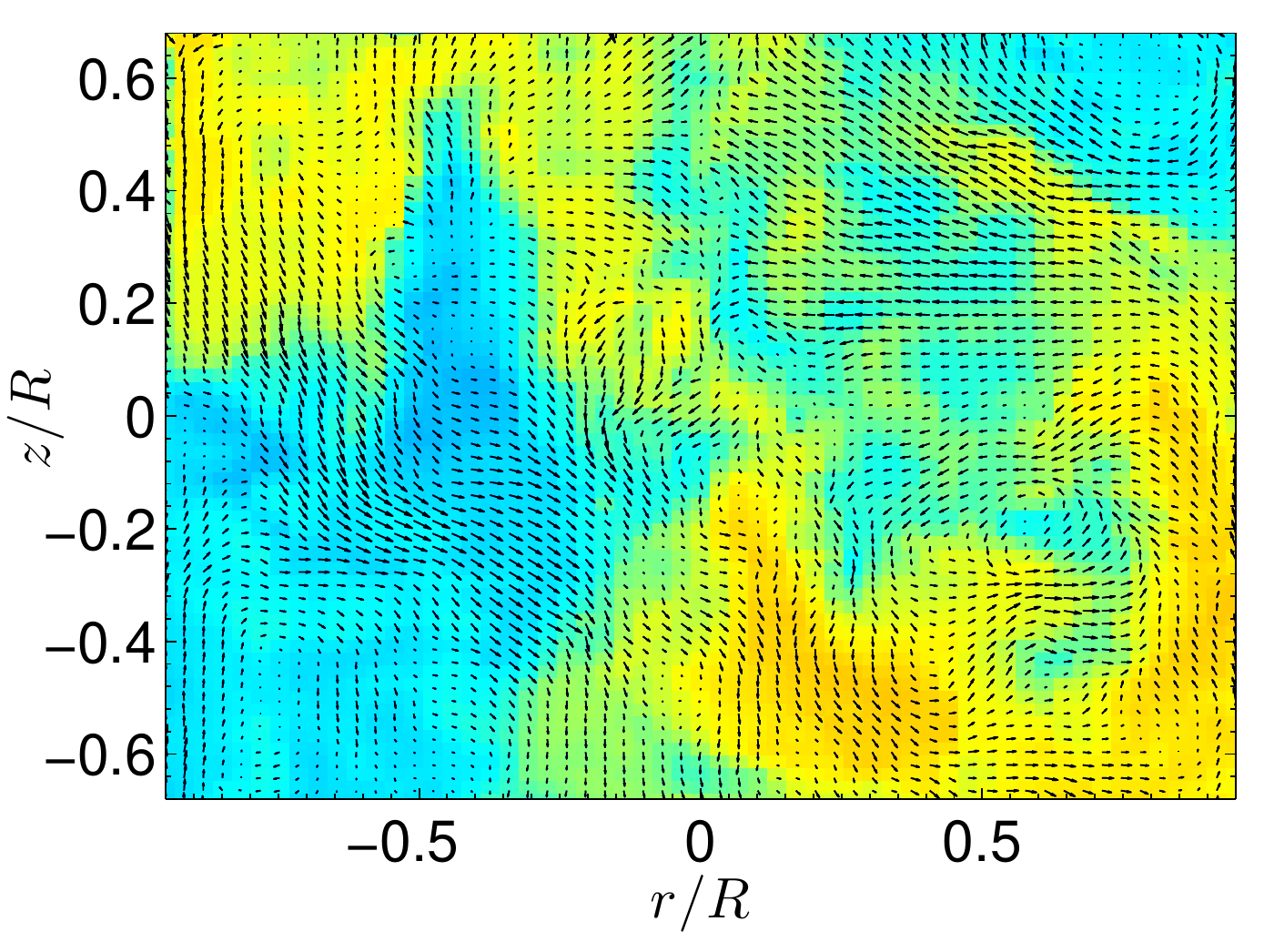} \\
  \includegraphics[width=0.99\columnwidth,trim = 0cm 0cm 0cm 0cm ]{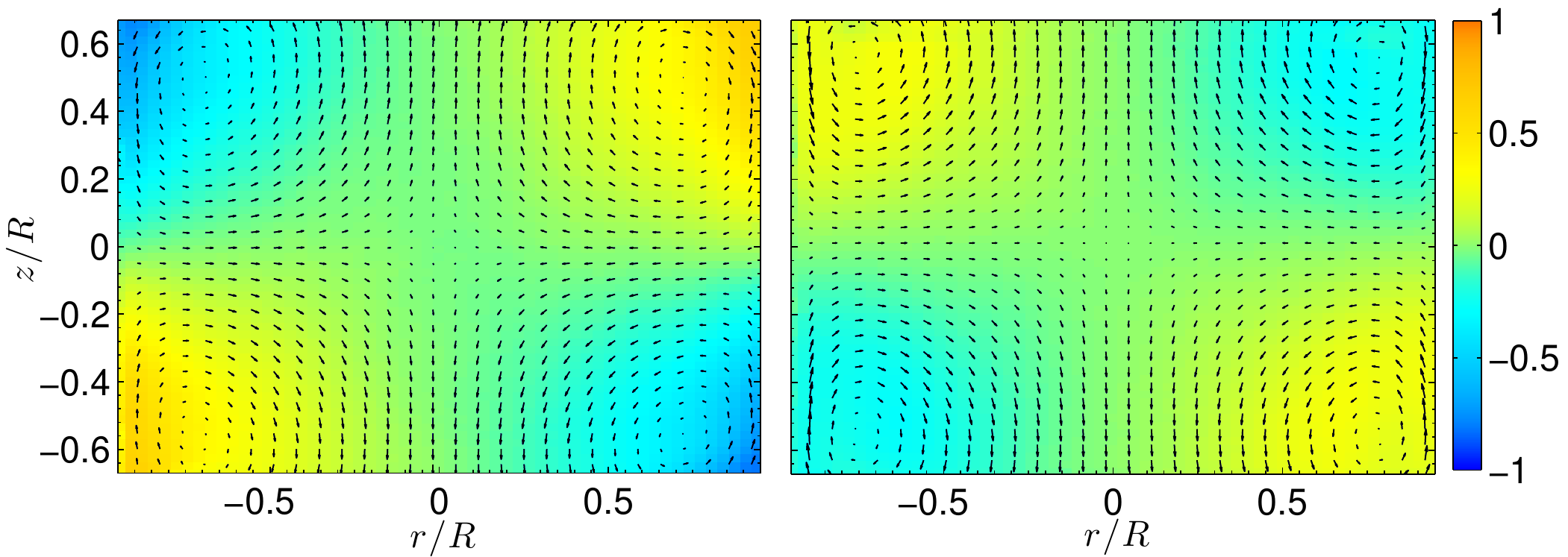}
\end{minipage}
\caption{Velocity fields reconstructed from  SPIV measurements at $Re \simeq 3 \times 10^5$ and $\theta=0$.  Top: instantaneous snapshot for TM87(+). Bottom : after time averaging over 600 snapshots,  for TM87(-) (left) and  TM87(+) (right). As the  velocity field is projected on a meridional plane that includes the rotation axis, the left part of the fields here  corresponds to ($u_r,-u_\phi,u_z)$) at $\phi=\pi$, while the right part corresponds to ($u_r,u_\phi,u_z)$) at $\phi=0$.
}
\label{fig:VI_instvw}
\end{figure}
As an example, Figure \ref{fig:VI_instvw} shows the $(r,z)$ dependence of the three components of a three-dimensional velocity field $(u_r,u_\phi,u_z)$ reconstructed from a PIV measurement at Reynolds number approximately equal to $3 \times 10^5$.  Although the SPIV system does not allow us to analyze in details the azimuthal dependence of the velocity fields, it is clear from Figure \ref{fig:VI_instvw} that the instantaneous velocity field is not axially symmetric, as it is not symmetric with respect to the transformation $r \to -r$.  Axial symmetry exists though at the level of the time averaged velocity field (and more generally for any quantity derived from it). At a coarse level of description, the topology of the average  velocity fields is simple  and appears to bear some kind of universality. It either consists of a two-cell state \footnote{Since the averaged flow is meant to be axially symmetric, the description refers to the flow observed in either the right or left half of the PIV window.} that is  symmetric with respect to the equatorial axis, a two-cell state with broken symmetry or a one-cell state.  

\subsubsection{A more quantitative description.}
\begin{figure}[!htb]
\centering
\begin{minipage}{0.99\columnwidth}
 \centering
 \includegraphics[width=0.39\columnwidth]{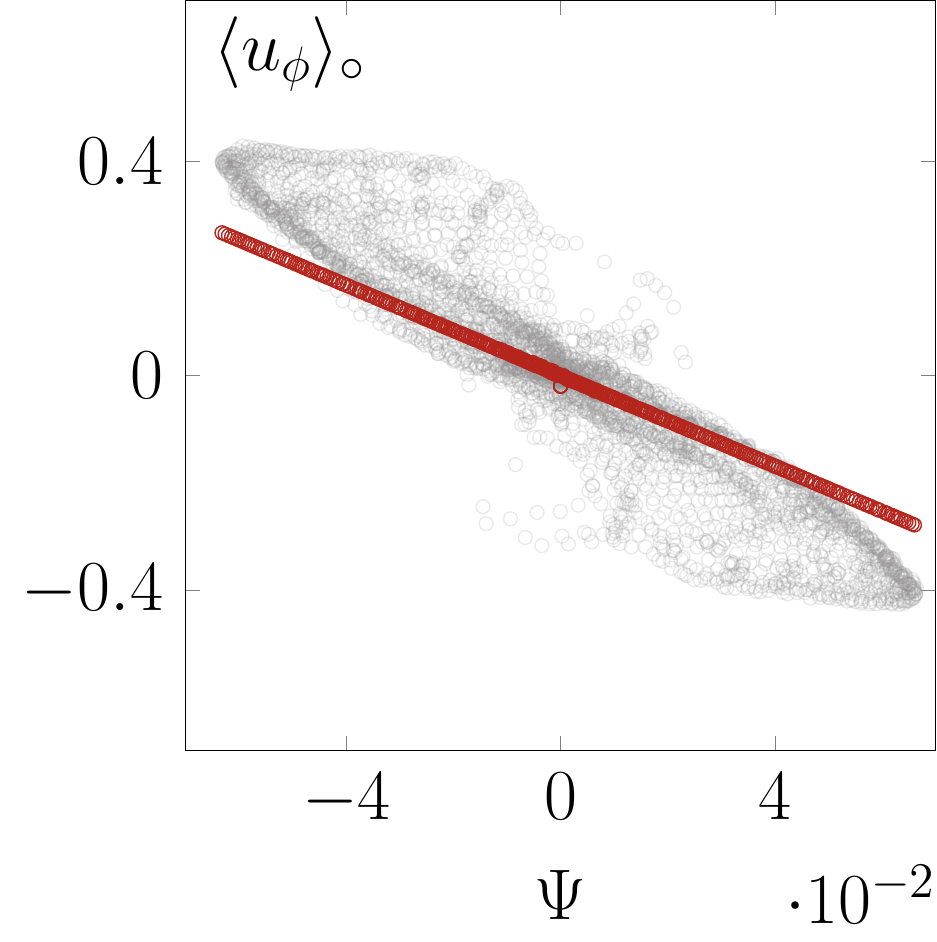}
 \includegraphics[width=0.1\columnwidth]{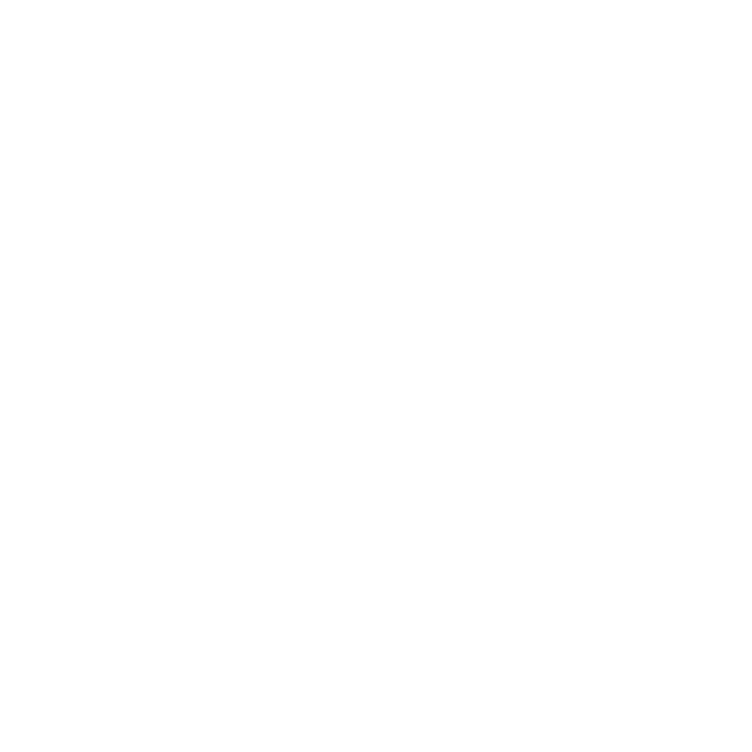}
 \includegraphics[width=0.37\columnwidth]{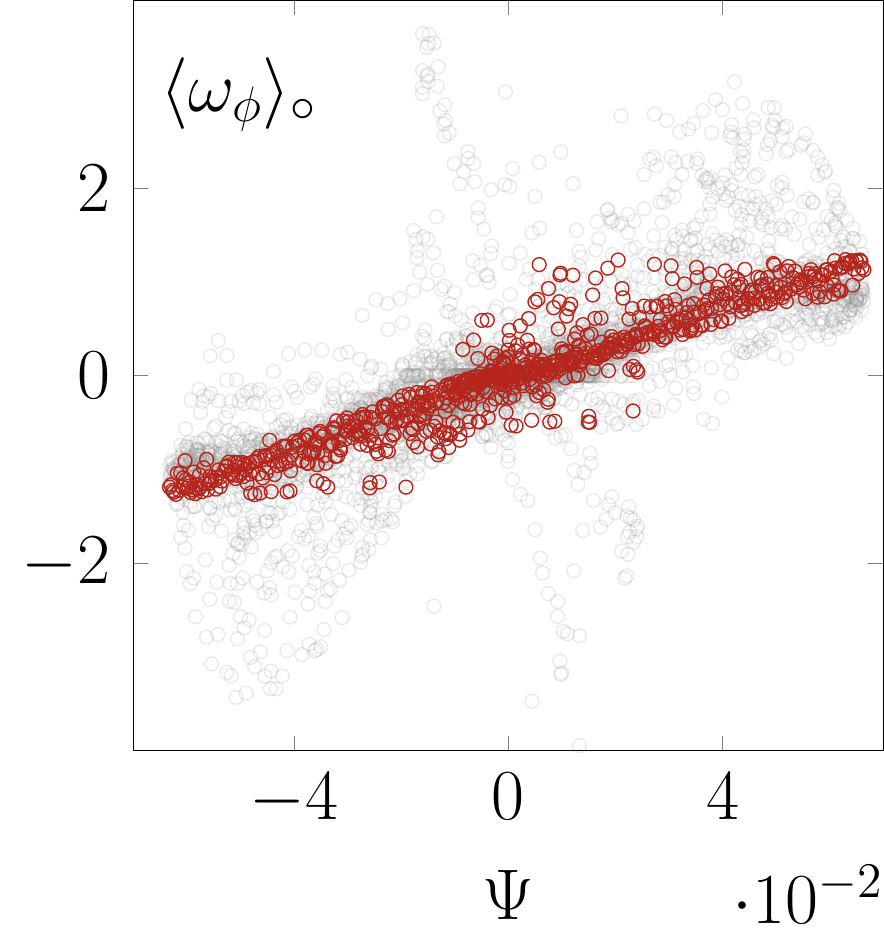}
 \end{minipage}
\caption{Scatter plots obtained in TM73(+) at $Re=10^5$ for  $(\theta,\gamma)=(0,0)$. The light grey dots are the data, the opaque red dots are the fits obtained from Equation (\ref{eq:axisteadylin}). Left: $\avc{u_\phi}$ as a function of $\Psi$. Right: $\avc{\omega_\phi}$ as a function of $\Psi$.}
\label{fig:ExampleFit}
\end{figure}

\begin{figure}[!htb]
\centering
\includegraphics[width=0.99\columnwidth]{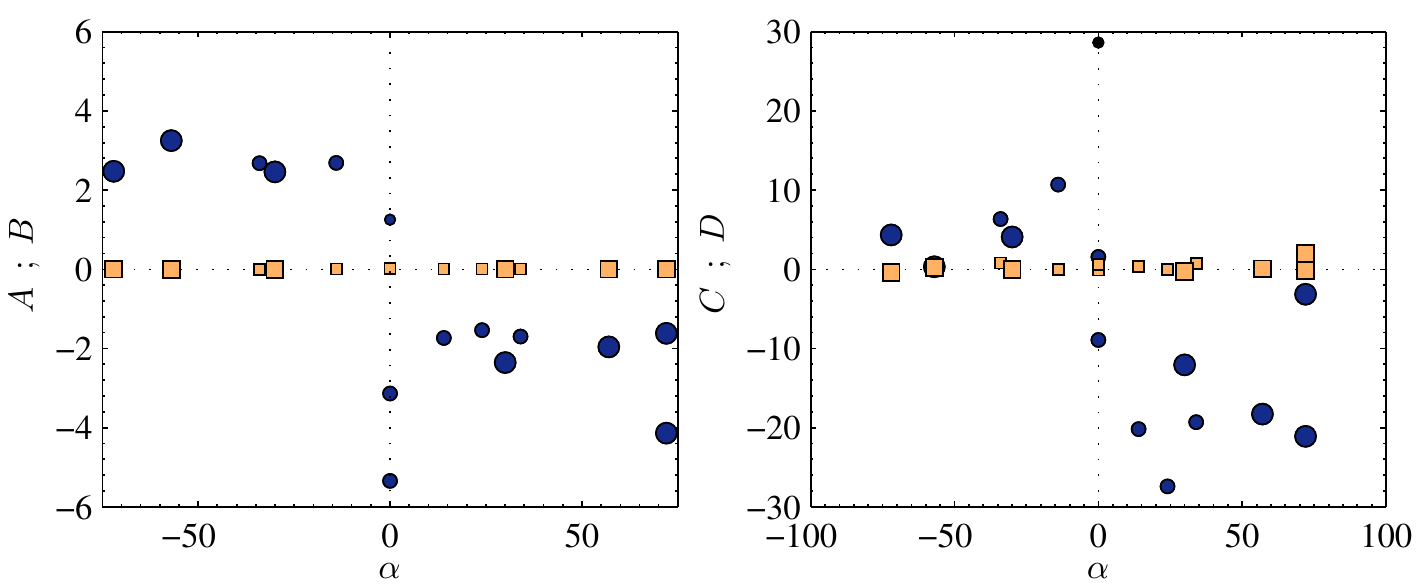}
\caption{The constant $A,B,C,D$ as defined by Equation (\ref{eq:axisteadylin}) to characterize the VK topologies . Left: $A$ (orange squares) and $B$ (blue circles) versus the angle $\alpha$; Right: $C$ (orange squares) and $D$ (blue circles) versus $\alpha$. The size of the symbol is proportional to the impeller's radius  $R_t=0.925$ (big), $R_t=0.75$ (medium) or $R_t=0.5$ (small).} 
\label{fig:ABCDVaryingAlpha}
\end{figure}

As first observed in \cite{monchaux2006steady}, the axial symmetry of the averaged 
states is not just an artifact of the averaging procedure, it can also be used as a natural guideline to describe the topology of the steady states.  As the average flow inside the tank is divergence free and axially symmetric (\emph{viz.} symmetric with respect to any azimuthal change), a Helmholtz decomposion can be used to write  the averaged velocity field as 
\begin{equation}
  \avc{\mathbf u} = \avc{u_\phi} \mathbf e_\phi + \dfrac{1}{r} \nabla(r \Psi) \times \mathbf e_\phi.
 \end{equation}
Independently of the underlying dynamics, the azimutal component of the vorticity $\avc{\omega_\phi} =( \nabla \times \avc{ \mathbf u}) \cdot \mathbf e_\phi$  is then related to the stream function through $\avc{\omega_\phi} =  -\partial_{zz} \Psi - \partial_r r^{-1}\partial_rr\Psi= - \mL(\Psi)$. The knowledge of $(\avc{u_\phi},\avc{\omega_\phi})$ is then sufficient to reconstruct the three dimensional averaged velocity field $\avc{\mathbf u}$. Using such a decomposition of the velocity field, Monchaux et al \cite{monchaux2006steady} evidenced that the axially symmetric averaged velocity fields observed in VK set-ups were peculiar steady solutions of the \emph{Euler} axially symmetric equations, at least in a region far from the propellers and the boundaries. The  Euler axially symmetric equations  are derived from the 3D (incompressible) Euler equations by considering the dynamics of a 3D velocity field whose cylindrical components do not depend on the azimuthal coordinate, and depend on $r$ and $z$ only ---\, see for example \cite{szeri1988nonlinear,leprovost2006dynamics} and Section \ref{sec:inviscid} below. Steady states of the axially symmetric Euler equations are obtained whenever the toroidal field $r u_\phi$, the poloidal field $\omega_\phi/r$ and the reduced stream function $\psi =r\Psi$ are related through relations of the kind  \cite{szeri1988nonlinear,leprovost2006dynamics} : 
\begin{equation}
 ru_\phi = F(\psi) \tieq{and} \omega_\phi/r-FF^\prime(\psi)/r^2 = G(\psi) \tieq{for any function $F$ and $G$.} 
\label{eq:axisteady}
\end{equation}
Using TM60($\pm$) propellers for a wide range of rotation numbers, scatter plots of both  the toroidal field $\avc{r u_\phi}$ and the poloidal field $\avc{\omega_\phi/r}$ against the reduced  stream function $\psi =r\Psi$ showed clear  functional relationship between those quantities \cite{monchaux2006steady,monchaux2007mecanique}.  To understand the general trends of the topologies, it is enough to choose  linear functions for $F$ and $G$. In this linear approximation, the VK topologies are  characterized by four constant numbers, say $A$, $B$, $C$ and $D$, defined by :
\footnote{To obtain (\ref{eq:axisteadylin}) from (\ref{eq:axisteady}), take $F(\psi) = B\psi+A$ and $G(\psi) = (C+DA/B) + D\psi$.} 
\begin{equation}
 \avc{u_\phi} = B\Psi +A/r \tieq{and} \avc{\omega_\phi}= (B+Dr^2)u_\phi +Cr.
\label{eq:axisteadylin}
\end{equation}
$A$, $B$, $C$, and $D$ are computed using least square fits from the scatter plots of $\avc{u_\phi}$ and $\avc{\omega_\phi}$ against $\Psi$. An example is shown in Figure \ref{fig:ExampleFit}, obtained for a large Reynolds number at $(\theta,\gamma)=(0,0)$. In such a case, the fit is rather good. As already noticed in \cite{monchaux2006steady}, the fit deteriorates at lower Reynolds numbers and when the rotation number is too high (say $\vert\theta\vert \gtrsim 0.3$). Still,  Equation (\ref{eq:axisteadylin})  provides a general framework for the interpretation of the data. 
The fitting procedure was carried with various propellers at large Reynolds number  in the symmetric state ($\gamma=0$ and $\theta=0$) with LDV data, that provide lower resolution representations of the mean flow.
The resulting values for $A$, $B$, $C$ and $D$ as a function of the propeller's radius and angle are provided in Figure (\ref{fig:ABCDVaryingAlpha}). Because of the measurement technique, these fits are less accurate than with the PIV data. The LDV-measured values of $A$, $B$, $C$ and $D$ should therefore be here used to observe trends rather than providing quantitative values.  We observe that in all cases, both $A$ and $C$ are vanishing. The fact that  $A =0$ is compatible with $u_\phi$ being finite at $r=0$, while the fact that $C=0$ is a consequence of the symmetry of the basic state at $\theta=0$. The value of $B$ depends mostly on the propeller's angle $\alpha$, being positive when the angle is negative and negative otherwise. The  absolute value of $B$ remains fairly constant in between $3$ and $4$, regardless of the angle ---\, except for the $\alpha=0$ case. 
At negative $\alpha$, the value of $D$ is rather low, and close to $0$. Increasing $\alpha>0$ yields a linear decrease for $D$, from $0$ to $-20$ (at $\alpha=72^\circ$). At this value of $\alpha$  though,  some indications exist, that a second branch of solutions can be found, with $D=0$. This has been confirmed by a study of the variations of our parameters with the Reynolds number, using PIV data, at $\theta=0$  in the symmetric state. With these better resolved data (not shown),  we found that the coefficient $D$ displays a clear bi-modal behavior, with two branches of solution :  One extends around $D=0$, and the other decreases linearly with $\log Re$. A closer look at the values of $D$ for the TM60/87(+) propellers indicates that the negative branch of solutions corresponds to a branch that connects continuously from a two-cell to a one-cell solution. As for the other coefficients, we found that $B$ is rather insensitive to the Reynolds number while both the $A$ and $C$ coefficients remain zero, at any Reynolds number, in agreement with the previously described regularity and symmetry properties. 

\subsection{Transitions between the various topologies}
\begin{figure}[!htb]
\centering
 \begin{minipage}{0.99\columnwidth}
 \includegraphics[width=0.99\columnwidth]{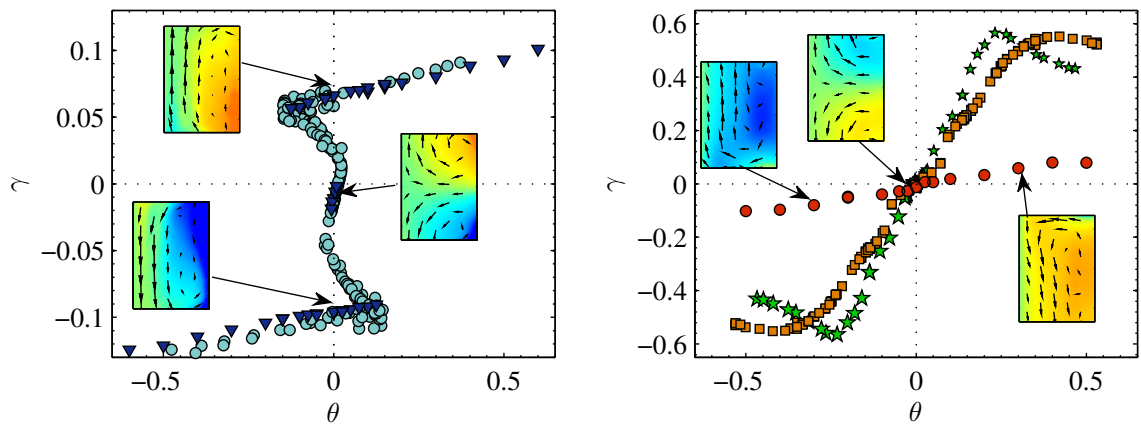}
 \end{minipage}
\caption{Variation of the topology of the stationary state, as a function of the forcing type. The symbols trace the torque asymmetry $\gamma$ versus the  averaged rotation number $ \theta$. Left: for TM87(-) at constant torque (circles) and constant speed forcing (triangles). Right: for TM87(+) (red circles) , TM73(+) (yellow squares) and TM73(-) (green stars) at constant speed forcing.  In insert, the corresponding topologies of the velocity field are shown.}
\label{GamavsThetaTorque}
\end{figure}

When the forcing is fully symmetric ($\theta=0$ and $\gamma=0$), all the impellers that we have tested yield  a symmetric  two-cell state similar to  the one depicted in Figure \ref{fig:VI_instvw}. 
Similarly,  when the forcing is clearly non-symmetric ($| \gamma | \simeq 1$ or $| \theta | \simeq 1$), the fluid is globally in  rotation and the  average  state is  one-cell.
Yet, the nature of the transition between the symmetric and the non-symmetric states does strongly depend on the geometry of the impellers. On the one hand,  the use of low curvature impellers ($\alpha \gtrsim -30^{\circ}$) yields a continuous transition that occurs via a sequence of increasingly non-symmetric two-cell states, with one cell becoming larger at the expense of the other. The sharpness of the transition can be characterized throughout the use of susceptibility coefficients (see Section \ref{sec:phasetransition}), analogous to the magnetic susceptibilities in the theory of ferro-magnetism.  Cortet et al \cite{cortet2010experimental,Cortet2011} observed that those susceptibilities diverge at a finite turbulent Reynolds number $(Re\simeq 4\times 10^4)$, a feature clearly reminiscent of second-order phase transition in statistical physics.
On the other hand, the use of  high curvature impellers ($\alpha \lesssim-30^{\circ}$) yields an abrupt change, that  gives rise to
multi-stability between the two-cell symmetric state, and one of the two one-cell states (symmetric to each other with respect to the equatorial axis) \cite{ravelet2005bifurcations}. As a result, a hysteresis cycle for $\gamma$ is described when  the rotation number $\theta$ is used as a control parameter, and cycled  from $1$ to $-1$ and back, over a given time scale. Increasing the curvature of the blades increase both  the width  and the height  of the hysteresis cycle that the system describes in the $\theta-\gamma$ plane\cite{ravelet2005bifurcations}. When the torque number $\gamma$ is used as a control parameter (\emph{i.e.} when the system is forced at constant torque rather than at constant velocity), the hysteresis cycle is regularized \cite{saint2013zero,saint2013forcing} ---\, see Figure \ref{GamavsThetaTorque}.  
Because of the apparent simplicity and universality of the steady states, there is  a good hope that a global understanding of the steady states can be provided through general arguments based on symmetries and conservation laws. This is precisely the outcome of statistical physics.  In the sequel, we try to explain the topology of the axially symmetric mean velocity field and explain their stability as a function of the control parameters $Re,\theta,\gamma, \alpha$, using some tools borrowed from   statistical physics.\\

\section{Insights from inviscid  theories}
\label{sec:inviscid}
Statistical theories of turbulent flows have so far only be conducted in the ideal case of Euler equations with symmetries \cite{robert1991statistical,miller1992statistical,Turkington2001,weichman2001statistical,chavanis2008statistical,naso2010statistical2,bouchet2010invariant,thalabard2013statistical}. In this section, we summarize ideas and the outcome of the statistical theories based on  the axially symmetric Euler equations. Most of the technical details are pushed to appendix, in order to focus on the predictions that those theories lead to. We then use the corresponding results as a guideline to understand the topologies of the flow inside the VK set-up.

\subsection{The axially symmetric perfect fluid.}
Inside an axially symmetric domain $\mD$, an axially symmetric fluid is described in terms of a 3D velocity field ${\bf u}$ , whose components in cylindrical coordinates $(u_r,u_\phi,u_z)$ do not depend on the azimuthal coordinate $\phi$ but on $r$ and $z$ only. The evolution equation of a perfect axially symmetric fluid is therefore obtained by setting $\partial_\phi = 0$ in the incompressible Euler equations 
\begin{equation}
 \partial_t {\mathbf u} + {\mathbf u} \cdot \nabla {\mathbf u} +\nabla p  = 0 \sieq \text{and} \sieq \nabla \cdot {\mathbf u} =0.
\label{eq:Euler}
\end{equation}
Rather than prescribing the 3D velocity field, a convenient description of an axially symmetric flow can be achieved in terms of only two  fields : \emph{i)} the azimuthal velocity $u_\phi$ (later named ``toroidal field''), and \emph{ii)} the azimuthal vorticity field $\omega_\phi = (\nabla \times {\bf u}) \cdot {\mathbf e}_\phi$ (alternatively named ``poloidal field'').
The entire 3D velocity field can then be reconstructed 
by using the incompressibility condition ($\nabla \cdot {\bf u} =0$) that allows the following Helmholtz decomposition for ${\bf u }$: 
\begin{equation}
{\bf u }= u_\phi{\mathbf e}_\phi + \dfrac{1}{r}\nabla (r \Psi) \times {\mathbf e}_\phi,
\end{equation}
where $\Psi$ is the  stream function, deduced from the  azimuthal vorticity by the relation :
%\begin{equation}
%\omega_\phi/r  = - \left(\dfrac{1}{r^2}\partial_{zz} + \dfrac{1}{r} \partial_{r}\dfrac{1}{r} %\partial_{r} \right) (r\Psi) = \mL (r\Psi). 
%\label{eq:xi-psi}
%\end{equation}
%
\begin{equation}
\omega_\phi  = - \left( \partial_{zz} +\partial_r \dfrac{1}{r} \partial_{r} r \right) \Psi = -\mL (\Psi). 
\label{eq:xi-psi}
\end{equation}
We also define $\psi = r\Psi$ the reduced stream function.
To invert the differential operator $\mL$, one needs to work with specified boundary conditions. We will assume here that $\psi$ is vanishing at the boundaries, a condition that comes for the impenetrability condition for the velocity field on the outer walls. 
Finally, for a cylindrical domain $\mD$ with height $2h$ and radius $R$, we will later use the short-hand notation $\int_\mD \equiv (1/(hR^2))\int_0^Rr\d r\int_{-h}^{h} \d z$ later. 
This  simple geometry will serve as a guideline for the statistical theories  described in the remainder of the section. 

\subsection{Analogy with a Ginzburg-Landau theory}
A basic input of the statistical physics of axially symmetric flows is the existence of conserved global quantities  in the inviscid, unforced limit ---\, at least provided that the perfect fluid can be considered  to remain ``sufficiently regular''.  
For instance, it is well known that the Euler dynamics (\ref{eq:Euler}) preserve  both the kinetic energy $E=1/2\int_\mD {\mathbf u}^2$  and the Helicity $H=\int_\mD {\mathbf u}\cdot (\nabla\times {\mathbf u})$ \cite{frisch1996turbulence}. In terms of  $u_\phi$ and $\omega_\phi$ these quantities read :  
\begin{equation}
 E = \dfrac{1}{2}\int_\mD  u_\phi^2 + \omega_\phi \Psi \sieq \text{and} \sieq  H = 2 \int_\mD  u_\phi \omega_\phi. 
 \label{eq:energyhelicity}
\end{equation}
\modif{
Axially symmetric velocity fields are three-dimensionnal fields which are invariant with respect to the azimuthal coordinate. This spatial continuous symmetry around the axis makes the axially symmetric Euler dynamics  further preserve two infinite families of global quantities, \emph{viz.} the toroidal Casimirs $C_f$ and the generalized helicities $H_g$ (see \cite{szeri1988nonlinear,leprovost2006dynamics}). This is a consequence of Noether's theorem and of the fact that the Euler dynamics is Hamiltonian \cite{morrison1998hamiltonian,holm1985nonlinear}. The additional invariants read} 
\begin{equation}
 C_f=\int_\mD f\left(ru_\phi \right) \tieq{and} H_g = \int_\mD g\left(ru_\phi\right)\omega_\phi/r  \text{~for any regular enough $f$ and $g$.}
\label{eq:axicasimirs}
\end{equation}
Those quantities allow to interprete thermodynamically  the linear description of the VK topologies provided by  Equations (\ref{eq:xi-psi}), (\ref{eq:axisteady}) and  (\ref{eq:axisteadylin}).
The idea is to use the invariants of Equation (\ref{eq:axicasimirs}) to build a ``free energy'' functional, that can  be thought of as an analog of  a Landau-Ginzburg functional \cite{monchaux2006steady,leprovost2006dynamics,naso2010statistical}. To stick to a linear level of description, one uses a quadratic form for the functions $f$  and a linear form for $g$, \emph{viz.}, $ f(x) = \alpha x^2 -\mu x$ and $g(x) = h x -\gamma$, with $\alpha$, $\mu$, $h$, $\gamma$ yet unprescribed scalars.   The free energy then  reads :   
%\begin{equation}
%  \mF[\sigma,\xi] = \int_\mD (\beta/2r^2+\alpha) \sigma(\br)^2 + h %\sigma(\br) \xi(\br)  + \beta \xi(\br)\psi(\br) /2  - \gamma \xi(\br) %- \mu \sigma(\br).  
%\label{eq:fluidginzburglandau}
%\end{equation}
%
\begin{equation}
  \mF[u_\phi,\omega_\phi] = \int_\mD (\beta/2+\alpha r^2) u_\phi^2 + h u_\phi \omega_\phi  + \beta \omega_\phi \Psi/2  - \gamma \omega_\phi/r - \mu r u_\phi.  
\label{eq:fluidginzburglandau}
\end{equation}
If the fluid inside the VK set-up was both perfectly axially symmetric and inviscid, then we could expect the local minimizers of $\mF$ to play a peculiar role.  Indeed, the reader familiar with dynamical systems may have already recognized that  the free energy (\ref{eq:fluidginzburglandau}) is an Arnold function relevant for axially symmetric perfect fluid, whose minima (if any) provide axially symmetric  profiles that are formally stable ---\, see \cite{holm1985nonlinear,szeri1988nonlinear} for more details about the stability of infinite dimensional dynamical systems and the stability  of axially symmetric perfect flows in particular \footnote{\modif{The  most general free energy for the axially symmetric dynamics is the functional  $E+C_f+H_g$ for prescribed $f$ and $g$. Equation \ref{eq:fluidginzburglandau} is a particular case.} }. The  critical points of $\mF$ are determined by the following  class of axially symmetric fields $(u_\phi^\star,\omega_\phi^\star)$ :
% 
%\begin{equation}
%\beta \sigma_c/r^2 + 2\alpha \sigma_c -+h \xi_c-\mu= 0  \tieq{and}  \beta \psi_c + h\sigma_c - \gamma=0.
%\label{eq:axiequilibrium}
%\end{equation}
\begin{equation}
(\beta + 2\alpha r^2) u_\phi^\star + h \omega_\phi^\star-\mu r = 0  \tieq{and}   hu_\phi^\star + \beta \Psi^\star - \gamma/r=0.
\label{eq:axiequilibrium}
\end{equation}
Setting $A=\gamma/h$, $B=-\beta/h$,  $C=\mu/h$ and $ D=-2\alpha/h$ , we exactly retrieve Equation (\ref{eq:axisteadylin}), which we previously used to characterize the VK topologies. This provides a clear connection between the steady states inside the VK tank and the steady states of the axially symmetric perfect fluid. However, it is easily shown that the fields $(u_\phi^\star,\omega_\phi^\star)$ that satisfy (\ref{eq:axiequilibrium}) \emph{do not} in general locally minimize the free energy $\mF$, unless $h=0$ (non-helical case). In this case,  positive values for $\beta$ and $\alpha$  ensure that  the $(u_\phi^\star,\omega_\phi^\star)$'s are indeed minimizers of $\mF$. 
%\footnote{In fact, a more general condition is $\alpha > -\int_\mD 1/(r^2)$.}
%
 If $h$ is non zero, the profiles that satisfy (\ref{eq:axiequilibrium}) are saddle points of the free energy and are therefore unstable with respect to any non-trivial perturbations.  Were we dealing with an inviscid axially symmetric fluid, would we therefore conclude  that  such ``meta-stable'' profiles could not be observed in the long term. It is however an experimental fact, that those profiles (with non zero ``h'') \emph{are} relevant to approximate those observed in a VK flow \cite{monchaux2006steady}.

\begin{figure}[!htb]
\centering
\includegraphics[width=0.74\columnwidth]{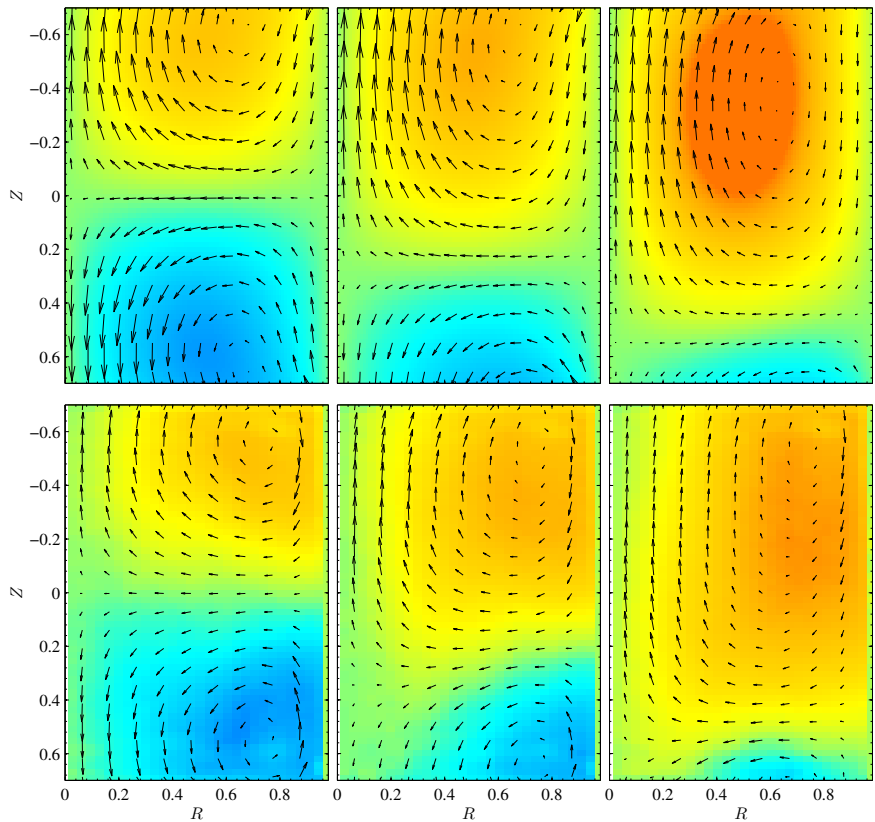}
\caption{Comparison between the velocity fields obtained in the VK experiment with TM73(+) propellers (top) and the Beltrami approximation (bottom) for ``VK boundary conditions'' at $\theta=0$ (left), $\theta=0.05$ (Middle)  and $\theta=0.09$ (right). The height and radius of the cylinders in the numerics are those of the real VK set-up, namely  $2H=1.8$ and $R=1$.} 
\label{fig:CompaFields}
\end{figure}

To connect the VK topologies with the meta-stable profiles of Equation  (\ref{eq:axiequilibrium}), some crude  identifications need to be made : the equilibrium fields $(u_\phi^\star,\omega_\phi^\star)$ of the inviscid theory with  the averaged PIV fields $(\avc{u_\phi},\avc{\omega_\phi})$, and the  axially symmetric domain $\mD$ with the PIV measurement domain $\Dpiv$. The VK analogs of the axially symmetric Casimirs  are  the Angular Momentum $I$, the Circulation $\Gamma$, the Helicity $H$, the toroidal energy $T$, and the poloidal energy $P$, defined as
\begin{equation}
\begin{split}
& I=\int_\Dpiv \avc{ru_\phi}\tieq{,} \Gamma = \int_\Dpiv \avc{\omega_\phi} \tieq{,} H=\int_\Dpiv \avc{u_\phi}\avc{\omega_\phi} \\
T= &(1/2)\int_\Dpiv {\avc{u_\phi}}^2  \tieq{and} P= (1/2)\int_\Dpiv {\avc{u_r}}^2+ {\avc{u_z}}^2.
\end{split}
\end{equation} 
Equations (\ref{eq:axiequilibrium}) and (\ref{eq:axisteadylin})  actually provide a good approximation of the steady states of the VK set-up..  Figure (\ref{fig:CompaFields}) shows the comparison between steady states obtained using TM73(+) propeller and solutions predicted by inviscid thermodynamics. To obtain it, we solved the equations (\ref{eq:axisteadylin}) with the SPIV-measured values of the constants $A$, $B$, $C$ and $D$ ---\, and appropriate boundary conditions \cite{saint2013zero}. To make the computation easier, we approximated the quantity  $B+Dr^2$ by a constant $K_{\text{eff}}$. We will later refer to this approximation as a ``Beltrami'' approximation.  
\footnote{ The equation to be solved is then simply obtained by combining the two lines of Equation (\ref{eq:axisteadylin}) ---\, or equivalently (\ref{eq:axiequilibrium}) \,--- and recalling that $\omega_\phi = - \mL \Psi$. It reads :   $-\mL \Psi=K_{\text{eff}}B\psi+AK_\text{eff}/r+Cr $ and is solved with $\Psi = 0$ at the boundaries ( ``vanishing boundary conditions'') or by prescribing that $u_\phi(r,z=\pm H)$ describe a solid rotation at frequencies $f_1$ and $f_2$ (``VK boundary conditions'' :  see \cite{saint2013zero} for more details). }

\subsection{The statistical mechanics perspective.}

\subsubsection{Analogy with a lattice model : a qualitative description.}
Since the topology of VK flows can be retrieved from a simple thermodynamic argument, one may wonder about the practical use of a full statistical theory.  However, we also noticed that the VK steady profiles did not match any axially symmetric genuine equilibrium, but rather a class of axially symmetric ``meta-stable'' equilibria. One aim of statistical mechanics is to understand this discrepancy , and  highlight important distinctions between VK flows and truly axially symmetric fluids. In particular, we will argue that although the VK profiles are in principles meta-equilibrium ones, they can still be interpreted as profiles that maximize a suitably-defined configuration entropy.

\begin{figure}[!hbt]
\centering
\includegraphics[width=0.49\textwidth]{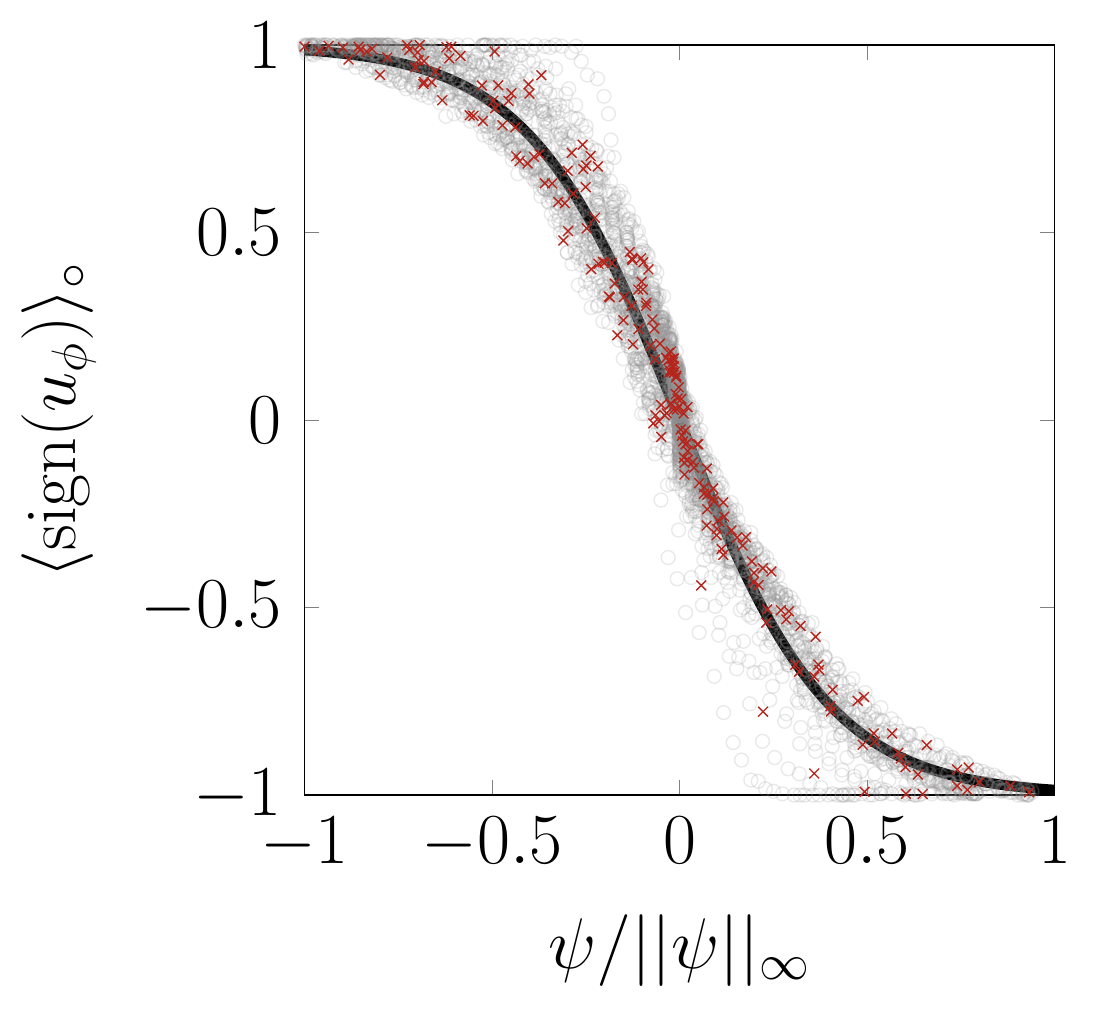}
\caption{The averaged  azimutal velocity sign scatter plotted against the reduced stream function $\psi = r\Psi$, at $\theta = 0$ , $\gamma=0$ and $Re \simeq 3 \times 10^4$ for TM60(+). $\psi$ has been normalized so that it ranges from $-1$ to $+1$. The light grey dots are the data, obtained after a time SPIV averaging. The red crosses are obtained by further averaging the signs on blocks of $4^2$ contiguous SPIV-lattice points. The black line indicates a $\tanh$ law.}
\label{fig:CurieWeissTurbulence}
\end{figure}

The statistical origin of the coarse-grained steady states can be qualitatively intuited, by looking at the averaged \emph{signs} of the azimuthal field for nearly symmetric forcing ($\gamma \simeq 0$ and  $\theta \simeq 0$).
In a simplified interpretation of the VK experiment, one may want to think about  the signs of the instantaneous azimuthal velocity field ( as measured on each position of  SPIV  grid ) as a ``spin'' that could take either a $+1$ or $-1$ value. In the sequel we drop the  precautional commas ``" and call it a spin. Then, each propeller could be thought of as a statistical reservoir of $'+'$ and $'-'$, ensuring the numbers of $'+'$ and $'-'$ to remain steady.  The scatter plot of the average azimuthal velocity  sign against the stream function  makes a hyperbolic tangent law emerge whatever the Reynolds number ---\, see Figure \ref{fig:CurieWeissTurbulence}. At a qualitative level, the $\tanh$ law  is reminiscent  of the $\sinh$ laws observed in decaying 2D turbulence \cite{montgomery1992relaxation},  or to some particular ``two-level discrete case'' found in the statistical theories developed for 2D inviscid flows in \cite{robert1991statistical,miller1990statistical,miller1992statistical}. It is also reminiscent of the mean-field closure equations that appear in the study of long-range lattice models of ferromagnetism. Consider for example the Curie-Weiss model, one of the simplest  lattice model of ferromagnet that can be treated analytically \cite{nishimori2001statistical}, and whose Hamiltonian reads: 
\begin{equation}
\mathcal H=-\dfrac{1}{2N}\sum_{i=1}^N\sum_{j \neq i}s_is_j - h\sum_{i=1}^N  s_i \tieq{with} s_i = \pm 1.
\end{equation}
 In the thermodynamic limit ($N \gg 1$), it is well-known that  the canonical free energy per site at temperature $\beta^{-1}$ of the Curie Weiss model can be written as an  infimum over all the possible values of the magnetization $\mu=(1/N)\sum_{i=1}^N  s_i$, namely  
\begin{equation}
 f(\beta)=\inf_\mu \left\lbrace \dfrac{\beta \mu^2}{2}- \log \cosh\beta(\mu+h) \right \rbrace.
 \end{equation}
The infimum is reached for the value of the magnetization that satisfies a self-consistent $\tanh$ law, namely $\mu = \tanh\beta (\mu+h)$. In the Curie-Weiss model, the $\tanh$ law emerges as a consequence of the interactions being long range and the mean-field approximation being exact. In fact, we show in the sequel that the  $\tanh$ relation between the stream function and the averaged azimuthal velocity spins in the VK set-up has some origin in this analogy with the Curie-Weiss model. %This is
%In the next section, we use the axi-symmetric Euler equations as a guide line to write down an effective statistical theory for the coherent states just described.
%
\subsubsection{Analogy with a lattice model : the Euler perspective.}
The statistical mechanics of inviscid fluids, as developed by Robert, Sommeria and Miller \cite{robert1991statistical,miller1990statistical} for the 2D case, aims to determine the  coarse-grained configuration  (if any)  that a perfect fluid is the most likely to adopt, were it described in terms of  an equilibrium ensemble, be it micro canonical, canonical, grand-canonical. The coarse-grained description can usually be  achieved  in terms of a ``macro-state probability field'' $p$. In the case of axially symmetric fluids, the latter is defined  as   
\begin{equation}
p_\br(\sigma,\xi) = \text{Proba}(r u_\phi=\sigma \text{~and~}  r^{-1} \omega_\phi=\xi \text{~in the vicinity of~} \br)  
\label{eq:macrostateproba}
\end{equation}
As the number of degrees of freedom of the underlying axially-symmetric Euler equations is formally infinite,  the statistical ensembles and corresponding macro-state probability fields need to be defined through an appropriate discretization and coarse graining, whose details are here omitted ---\, see \cite{thalabard2013statistical} for the axially symmetric case and \cite{miller1992statistical,potters2013sampling} for the  2D case. This construction provides a useful analogy with standard lattice models of Ferro-magnetism. Indeed, the statistical ensemble that we wish to compute can now be thought of as the thermodynamic spin-wave limit  of a finite-size bi-dimensional lattice model of $N^2$ interacting spins $\bS$, whose Hamiltonian is prescribed by the specific shape of the inviscid dynamical invariants given by Equation (\ref{eq:axicasimirs}).
In the axially symmetric case, the spins $\bS_{ij}=(\sigma_{ij},\xi_{ij})$ are two-degrees-of-freedom objects  that represent the micro-scale value of the pair $(ru_\phi,\omega_\phi/r)$ at position $\br_{ij}$ ($1\le i,j \le N$). 
The analogy is clearer in the particular situation where the $\sigma_{ij}$'s can take only two values say $\sigma_\pm = \pm 1$ \footnote{Although particular, the case is easily extended to the general case where the $\sigma_{ij}$'s span a continuous range of values (see \ref{app:axiequilibria} for more details).}, (which would be the natural assumption if we wanted for example to study the statistics of the signs of the $\sigma_{ij}'s$ only). In this case, the appropriate micro-canonical lattice model interaction Hamiltonian per spin is 
\begin{equation}
 \mathcal H/N^2  = \dfrac{1}{N^4}\sum_{i,j,k,l} J_{kl}^{ij}\xi_{ij}\xi_{kl} \text{~~with   $J$ a discretization of -$\mathcal L^{-1}$},
\end{equation}
%\footnote{ou l'ecrire en fonction de psi}
subject to the constraint that both the toroidal magnetization $I=(1/N^2)\sum_{i,j} \sigma_{ij}$ and  the poloidal conditional magnetizations $\Gamma_\pm= (1/N^2)\sum_{i,j:\sigma_{ij}=\pm} \xi_{ij}$ remain constant.
It however turns out that the statistical ensembles  related to this ``axially-symmetric lattice model'' are  ill-defined in the thermodynamic limit \cite{leprovost2006dynamics,thalabard2013statistical}. The problem comes from the  poloidal degrees of freedoms, namely the $\xi_{ij}$'s, which are not sufficiently constrained  by the prescription that both the energy and the conditional magnetizations remain finite in the limit  $N \to \infty$. As a consequence, an ultra-violet catastrophe occurs : unless the averaged poloidal field is uniformly zero, the most probable configurations are those for which most of the $\xi_{ij}$'s are infinite and mask the  coherent structures. %\simon{ici, on peut citer jordan turkington weichman pour dire que la situation ressemble a ce qui se passe dans les lattice model de MHD so far derived}
Two problems are now apparent. The first one is theoretical. It relates to the existence or not of equilibrium measures for the axially symmetric Euler equations. The second problem is more practical : how can a ill-defined statistical theory possibly describe the topology of the flows observed inside the VK set-up? 
As previously emphasized, the present paper does not deal with the axially symmetric Euler equations but with VK turbulence. Therefore, the first question is clearly far beyond our present concerns. We now describe a strategy to deal with the second question.

\subsubsection{A phenomenological treatment  for the vorticity fluctuations.} 
\paragraph{The main idea.}
\modif{Over the past decade,  many statistical theories have been brought up to describe the statistical equilibria of the ideal axially-symmetric perfect fluid  \cite{mohseni2001statistical,lim2003coherent,leprovost2006dynamics,naso2010statistical,naso2010statistical2,thalabard2013statistical}. Those theories yield different outcomes. The reason stems from the existence of a so-called ``Ultra-Violet'' (UV) catastrophe in the ideal theory for the axisymmetric fluids.  In other words : due to the special nature of the inviscid invariants, a full equilibrium statistical theory for the axially symmetric fluid cannot predict both a finite non zero energy and a non trivial large-scale flow. 
This degeneracy originates from the non-zero vorticity stretching term, that allows any initial vorticity field to grow unbounded in the limit of a vanishing viscosity.
In order to write down a self-consistent and non-degenerate statistical theory, one therefore needs to \emph{model} the statistics of the small scales. Different \emph{Ansatz} lead to different outcomes that may or may not be relevant to describe VK flows. In order to select the suitable Ansatz, we first review the gist of the theories in the next paragraph. Their outcomes are summarized in Table \ref{Table:equilibria}. We then discuss the relevance of those theories to VK flows and define two statistical inverse temperatures.}

\begin{figure}[!htb]
\centering
 \includegraphics[width=0.99\columnwidth]{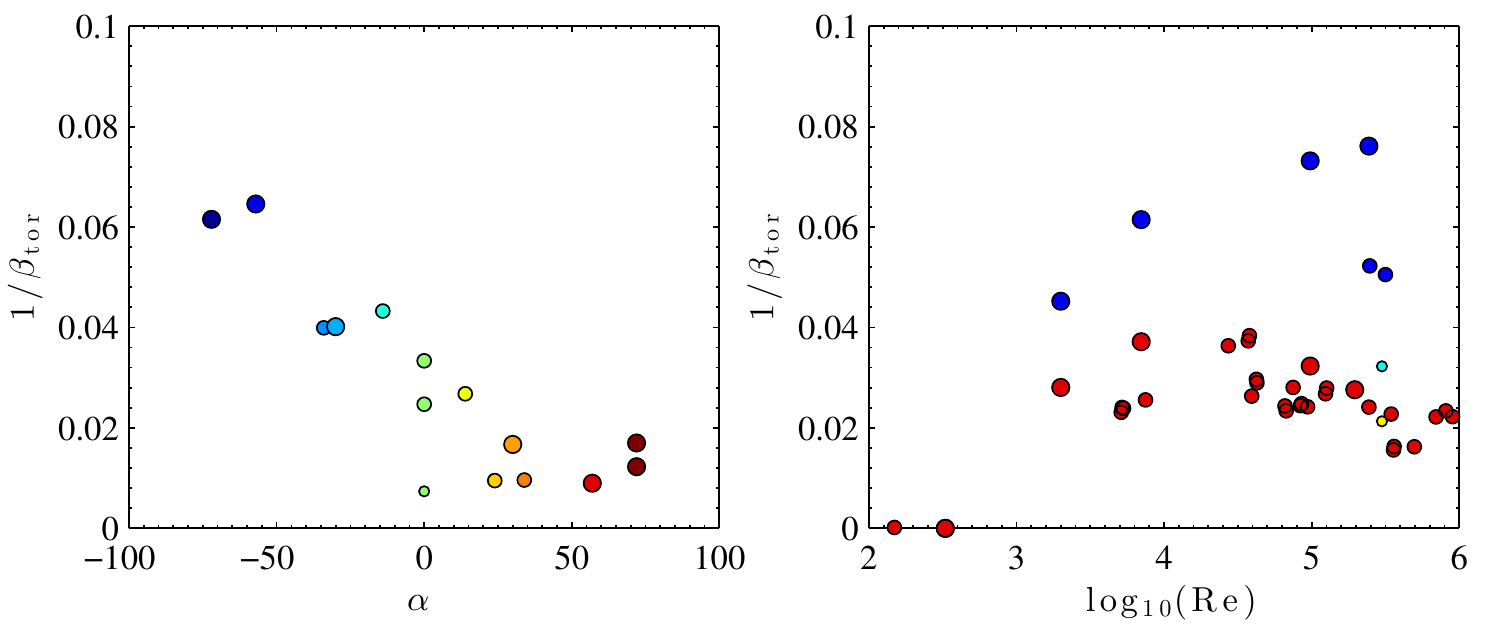}
\caption{The toroidal temperature $1/\beta_\tor$ as a function of the control parameters $\alpha$ (left) and $Re$ (right). The symbol color codes $\alpha$ (defined on the left panel).  The size of the symbol is proportional to the impeller's radius, $R_t=0.925$ (big), $R_t=0.75$ (medium) orl $R_t=0.5$ (small).} 
\label{fig:betapoltor}
\end{figure}

%
% At a practical level the questions raised are the following :  
%Are those outcomes also compatible with the observed VK averaged topologies ? Which ansatz is then the most satisfying one ? 
%For thoroughness, we describe the gist of the theories is the next paragraph, that can be omitted on first reading. The reader in a hurry can refer to Table \ref{Table:equilibria} to grasp the topologies yielded by the different theories.  }
%
\paragraph{The Gist of the ideal theories.}
As just mentioned, in common to all the works on axially-symmetric fluid equilibria is the use of additional phenomenological assumptions to control the fluctuations of the poloidal degrees of freedom. Ensemble averages are usually computed in terms of a most probable macro state probability field that maximizes a macro-state entropy, determined by  an appropriate use of Laplace's theorem, and the  saddle-point method. The idea is therefore to restrict the set of macro state probability fields over which the maximization is carried on.
References \cite{lim2003coherent,mohseni2001statistical} impose that the axially symmetric flow has a vanishing toroidal field, so that an extra-enstrophy constraint appears for the poloidal field; References \cite{leprovost2006dynamics,naso2010statistical,naso2010statistical2} assume that the poloidal field is non-fluctuating, \emph{viz.}, $\av{\omega_\phi^2} = \av{\omega_\phi}^2$ at any position $\br$ ; Reference \cite{thalabard2013contributions} suggests to freeze the poloidal degrees of freedom, \emph{viz} write the  macro-state probability field as  $p_\br(\sigma,\xi) = p_\br(\sigma) p_\br(\xi|\sigma)$ and prescribe the conditional probability distribution $p_\br(\xi|\sigma)$ ;   it was also suggested to impose a cut-off on the poloidal degrees of freedom, either used as a physical parameter \cite{leprovost2006dynamics}  or  as an intermediate regularization constraint \cite{thalabard2013statistical}. \\
Table \ref{Table:equilibria}  presents the different outcomes that the various ansatz lead to. Results are here summarized for two cases. In the ``Gaussian modeling'', only the quadratic and linear invariants enter the theory. In the two-level modeling, it is assumed that the toroidal degrees of freedom can only take the two values $\pm 1$.   Both cases give consistent results and are easily extended to the general case (see the appendix).  %However, they are here sufficient to provide us with some physical insights.  
Table \ref{Table:equilibria} shows that the different modelings yield different outcomes. This suggests that the poloidal fluctuations play a crucial role in VK mixing. \\
\renewcommand{\arraystretch}{1.45}
\begin{table*}[ht]
\centering
\resizebox{0.99\textwidth}{!}{
\begin{tabular}
{||c||c|cc||}%
\hline \hline
Name & Macrostate & Gaussian & Two-Level  \\
\hline 
\multirow{2}{*}{Frozen-$\br$} & \multirow{2}{*}{$\delta(\xi-\xi_0(\br)) p_\br(\sigma |\xi)$}& $\langle \sigma \rangle = \left(\mu - h\xi_0(\br) \right)/D(\br) $  & $\langle \sigma \rangle = \tanh(A+B\xi_0(\br))$\\
& & $\langle \xi \rangle = \xi_0(\br)   $& $\langle \xi \rangle = \xi_0(\br) $\\
%& & $\langle \sigma ^2 \rangle = \langle \sigma \rangle^2 + 1/(2\alpha(\br) )$  & \\
%& & $\langle \xi^2 \rangle = \langle \xi \rangle^2  $& \\
\hline 
\multirow{2}{*}{Frozen-$\sigma$} & \multirow{2}{*}{$p_\br(\sigma) \underbrace{p_\br(\xi | \sigma)}_{\text{prescribed}}$}& $\langle \sigma \rangle = (C- \beta A(\br) \psi (\br) )/D(\br) $  & $\langle \sigma \rangle = \tanh(A_1+B_1\psi(\br))$\\
& & $\langle \xi \rangle = A(\br) \langle \sigma  \rangle+ B  $& $\dsp \langle \xi \rangle = \dfrac{\xi_+\e^{A_++B_+\psi} + \xi_-\e^{A_-+B_-\psi}}{\e^{A_++B_+\psi}+ \e^{A_-+B_-\psi}}$\\
\hline
\multirow{3}{*}{Gaussian} & \multirow{2}{*}{$p_\br(\sigma,\xi)$} & $\langle \sigma \rangle = (A(\br)+ B(\br) \psi (\br) )/D(\br)$  &$\langle \sigma \rangle = \tanh(A_1+ B_1 \psi(\br) ) $ \\
& & $\langle \xi \rangle = (C(\br)+ F(\br) \psi (\br) )/D(\br) $& $\dsp \langle \xi \rangle = -\beta \psi/(4\nu) $\\
 &with $\langle \xi^2\rangle < +\infty$ & & $\dsp + \dfrac{1}{2\nu} \dfrac{g_+\e^{A_++B_+\psi} + g_-\e^{A_-+B_-\psi}}{e^{A_++B_+\psi} + e^{A_-+B_-\psi}}$\\
\hline
\multirow{2}{*}{Microcanonical} & \multirow{1}{*}{$p_\br(\sigma,\xi)$} & $\langle \sigma \rangle = A(\br)$  &$\langle \sigma \rangle = A_1(\br)$ \\
&with $\langle \xi^2\rangle \to +\infty$ &  $\langle \xi \rangle = B \psi (\br) + C $ & $\langle \xi \rangle = B_1 \psi (\br) + C_1$\\
\hline
\hline
\end{tabular}}
{\caption{Equilibria obtained using different ansatz for the azimuthal vorticity $\xi$. Greek letters denote Lagrange multipliers, Latin letters combinations of those ---\, that are different in each cell.  In the Gaussian modeling only the quadratic invariants are taken into account, while in the two-level theory it is assumed that the toroidal field can only take two values $\pm 1$. More details can be found in the appendix.} \label{Table:equilibria}}
\end{table*}
\renewcommand{\arraystretch}{1}
\paragraph{Discussion : Temperatures and VK mixing.}
If one indeed takes for granted that the axially symmetric Euler equations can be used as a guideline to understand the VK topologies ---\, which we remind is far from obvious but relies on experimental observations \,---, then those statistical theories shed a qualitative light on the mixing processes at stake inside the VK tank. 
On the one hand, the assumption that the  poloidal field is non fluctuating and fixed in space (the ``frozen-$\br$'' model) so that only toroidal degrees of freedom do indeed mix is not satisfying. This assumption yields only one of the two constitutive thermodynamic equations (\ref{eq:axiequilibrium}). It thus does not allow to self-consistently determine the averaged field  without a further ansatz about the poloidal profile. 
On the other hand, if one allows for too high a level of fluctuations in the theory (the ``micro-canonical theory'' ), one obtains a closed but very restricted class of profiles, \emph{viz.}, the ``$(h=0)$'' solutions of Equation (\ref{eq:axiequilibrium}). Those solutions correspond to cases where the coarse-grained toroidal field is completely decoupled from the stream function. Those solutions may therefore be formally relevant for an axially-symmetric perfect fluid,  but  are clearly not those observed inside the VK tank.  
Rather, the mixing inside the tank seems  to lie in between those two extreme cases. The ``frozen-$\sigma$'' theory provides such an example of mixing. The theory describes  poloidal degrees of freedom that cannot mix independently from the toroidal ones. Observe from Table \ref{Table:equilibria} that this assumption leads to a $\tanh$ relation between the coarse-grained toroidal field and the stream-function, as apparent in Figure \ref{fig:CurieWeissTurbulence}. \\
This interpretation of statistical mechanics  provides  a zero order approximation to the actual nature of the VK stirring.
The difference between the equilibrium steady states (the ``$h=0$'' case) and the non-equilibrium ones is here provided through a restriction of phase space. This idea is similar to the notion of restricted partition functions recently proposed by Herbert \cite{Herbert2013} to explore the nature of inverse cascades in helical turbulence. 
Finally note that our discussion was here restricted  to  predictions for the  averages of  first-order quantities. Obviously, a statistical theory based on an ``axi-symmetric ansatz'' fails to give an accurate and quantitative view about the fluctuations measured in VK turbulence, as those are clearly observed \emph{not} to be axially symmetric ---\, see Figure \ref{fig:VI_instvw}. It is however interesting to remark that the minimal ``frozen-$\sigma$'' theory does not prescribe any shape for the poloidal fluctuations. In particular, those are not prescribed to  be Gaussian, as one could in principle  expect from a statistical mean field theory that uses quadratic invariants as inputs. Because of the seemingly crucial role played by the fluctuations, it is natural to investigate in more details both the poloidal and the toroidal fluctuations inside the VK tank. To this end, we introduce two quantities, which can be thought of as two toroidal and poloidal inverse temperatures, \emph{viz.} $\beta_\tor$ and $\beta_\pol$. Those are  defined by
\begin{equation}
 \beta_\tor^{-1} = \int_\Dpiv \left(\avc{u_\phi^2}-\avc{u_\phi}^2 \right) \tieq{and} \beta_\pol^{-1}= \int_\Dpiv \left(\avc{\omega_\phi^2}-\avc{\omega_\phi}^2\right) 
\label{eq:VKtemperatures}
\end{equation}
These temperatures should depend on the experimental control parameters that can now be seen as analogous of "thermostats". In Figure \ref{fig:betapoltor}, we show for example the dependence of $1/\beta_\tor$ as a function of the angle $\alpha$ (left panel) and the Reynolds number  $Re$ (Right panel). It can be seen that the toroidal temperature increases with decreasing $\alpha$, and increases from zero past $Re\sim 10^3$ (the laminar/turbulent transition). The temperature peaks around $Re=40 000$ for TM60(+) impellers, and then saturates or slightly decreases.
Both the toroidal and the poloidal  temperature will allow us to describe VK topologies at any control parameter from a statistical perspective, that goes beyond the insight of the inviscid theories. 

\section{Statistical mechanics beyond the inviscid case : Analogy with the Curie-Weiss theory of Ferro-Magnetism}
\label{sec:phasetransition}
\begin{figure}
\centering
\includegraphics[width=0.79\columnwidth]{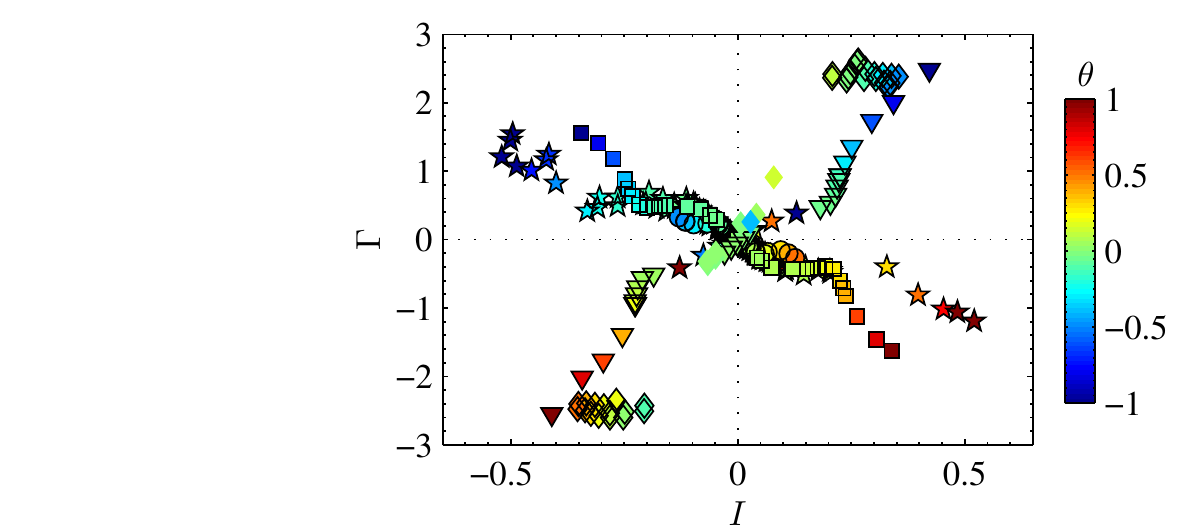}
 \caption{Magnetization $\bf M=(\Gamma,I)$ for various rotation numbers  $\theta$'s, and  different propellers : TM87 (+)(squares), TM87(-) (diamonds), TM73 (+)(stars)  and TM73 (-) (triangles). The color codes $\theta$.}
 \label{fig:spinM}
\end{figure}

%\subsection{Analogy with the  Curie-Weiss theory of Ferro-magnetism}
A major drawback of the theories described in the previous section is their intrinsic rooting on an inviscid description (\emph{i.e.} force-free, zero viscosity) of the VK flows. As such, they do not predict anything about finite Reynolds number effects. Besides, the presence of forcing and dissipation, as well as the lack of instantaneous axial-symmetry concur to destroy  the conservation of the global invariants on which the statistical description relies on. 
However, the broad description of the VK flow topologies in terms of an equilibrium statistical theory  allows for a thermodynamical interpretation of the transitions, in the spirit of the Curie-Weiss theory of Ferro-Magnetism. 
Here, we shall not specify in as much details as in the previous section the spin lattice model that we consider.
What we retain from the analogy is the following.  The VK flow can be seen as a lattice model, whose spins each have  two components : one linked with the toroidal velocity $u_\phi$, the other one linked with the toroidal vorticity $\omega_\phi$. Those two-component spins evolve under the action of both a thermostat and a symmetry breaking external field, that are provided by the four control parameters $Re$, $\gamma$, $\theta$, $\alpha$. The ability of the spin to orientate itself as a function of the forcing can be traced by the local ``magnetization vector'' $(ru_\phi(\br), \omega_\phi(\br))$, whose (space-time) average reads : 
\begin{equation}
{\bf M}= \left(I,\Gamma\right).
\label{eq:Magnetization}
\end{equation}
As illustrated in Figure \ref{fig:spinM}, different shapes for the propellers give rise to different behavior for  $ \bf M$, implying different ``preferred orientations'' for the spins. We think that the propensity of each spin to deviate from this orientation can be captured by the behavior of the two temperatures $\beta_\pol$ and $\beta_\tor$ based on toroidal vorticity and velocity fluctuations, see Equation (\ref{eq:VKtemperatures}) :  The more curved the propellers, the higher the temperatures.  In the same way, increasing the Reynolds number for a given shape increases the temperatures. 
In the spirit of statistical physics, the transition from a two-cell state towards a one-cell state can be thought of as a symmetry-breaking transition. It is then natural to introduce a susceptibility vector $\mathbf \chi$  as :
\begin{equation}
 {\mathbf \chi} = (\chi_I,\chi_\Gamma) = \left. \dfrac{\partial {\bf M} }{\partial \theta} \right|_{Re, \alpha}
\label{eq:susceptibilitydef}
\end{equation}
The complete analogy is summarized in Table \ref{Table:analog}. It allows for an interpretation of the  salient hydrodynamical observations previously observed in the VK set-up  in terms of phase transition and  critical exponents. With this analogy in mind, a new light can be shed on the finite-Reynolds number ``phase transition''  observed in the VK set-up by \cite{cortet2010experimental,Cortet2011}. The main signature of the phase transition is obtained by monitoring the behavior of the susceptibility $\chi_I(\theta=0)$, in order to characterize the situation of a weak symmetry breaking due to the external forcing.  In \cite{cortet2010experimental,Cortet2011}, the control parameter was taken to be the logarithm of the Reynolds number, in analogy with a definition formulated by Castaing in \cite{castaing1996temperature} for the  temperature of a turbulent flow. Here, we show that the phase transition can be identified and further characterized by the fluctuations of $u_\phi$ and $\omega_\phi$, which play the role of temperatures, as suggested by the inviscid statistical theory.

\begin{figure}[!htb]
\centering
 \includegraphics[width=0.99\columnwidth]{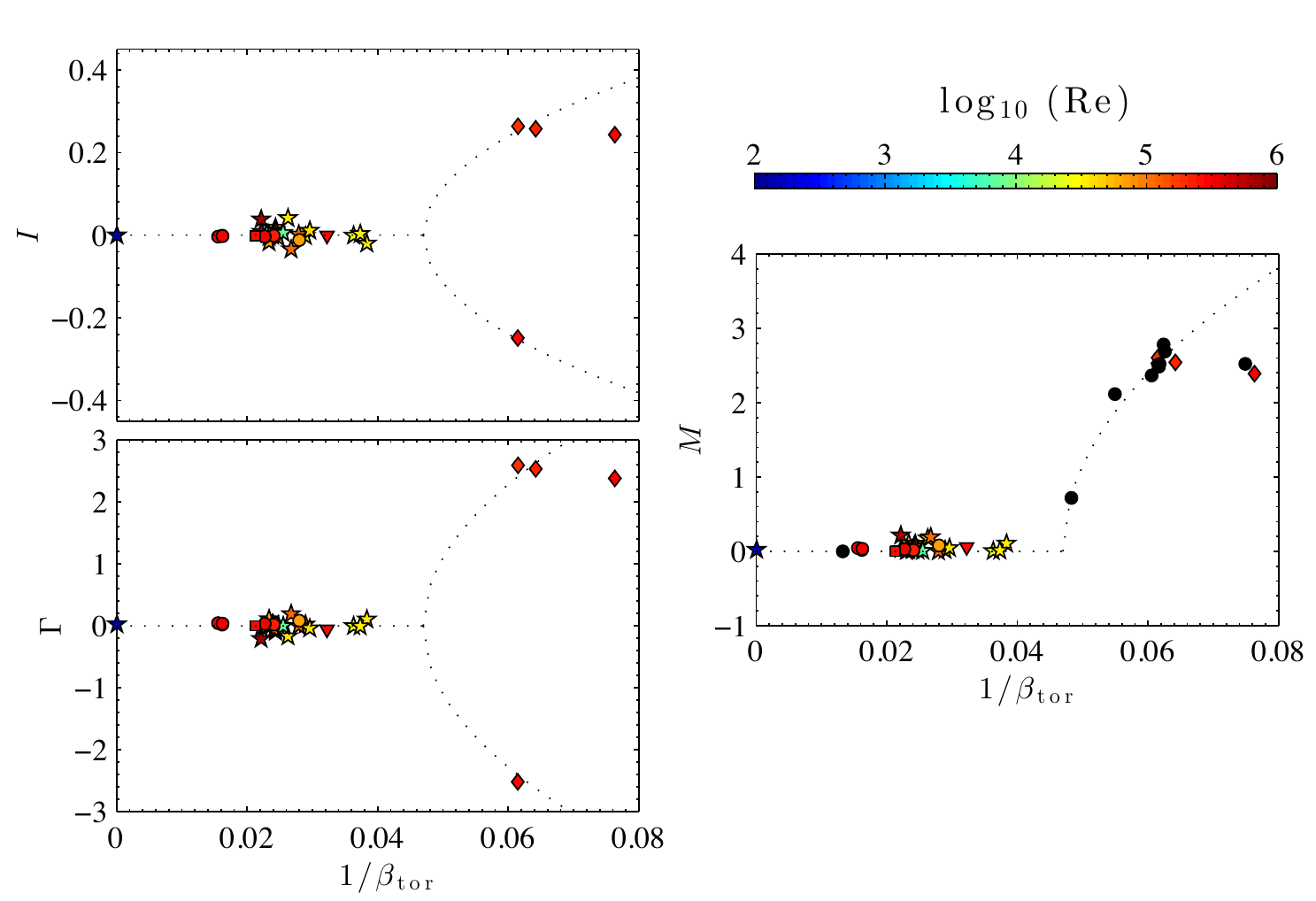}
\caption{Mean magnetization as a function of temperature. Left : 
 $I$ (top) and $\Gamma$ (bottom) as functions of $1/\beta_{\tor}$. 
Right : The magnetization $M= \sqrt{I^2+\Gamma^2}$ as a function of the toroidal temperature for different propellers (TM87(+)(circles) and (-) (diamond) and TM73 (+)(triangles)  and (-) (stars)) . In every case, the black dotted line indicates a fit $y=\pm a\sqrt{1/\beta-1/\beta^\star}$, with $1/\beta_\star=0.044\pm0.03$, and various $a$ : $a=2.1$ for $I$, $a=20$ for $\Gamma$ and $a=21$ for $M$. The black circles are magnetization estimates based on the height of the hysteresis cycle \protect\cite{saint2013zero} .%\simon{Ces points n apparaissent plus sur cette version de la figure. Me trompe-je ?}
}
\label{fig:Magnetization}
\end{figure}

\begin{figure}[!htb]
\centering
 \includegraphics[width=0.99\columnwidth]{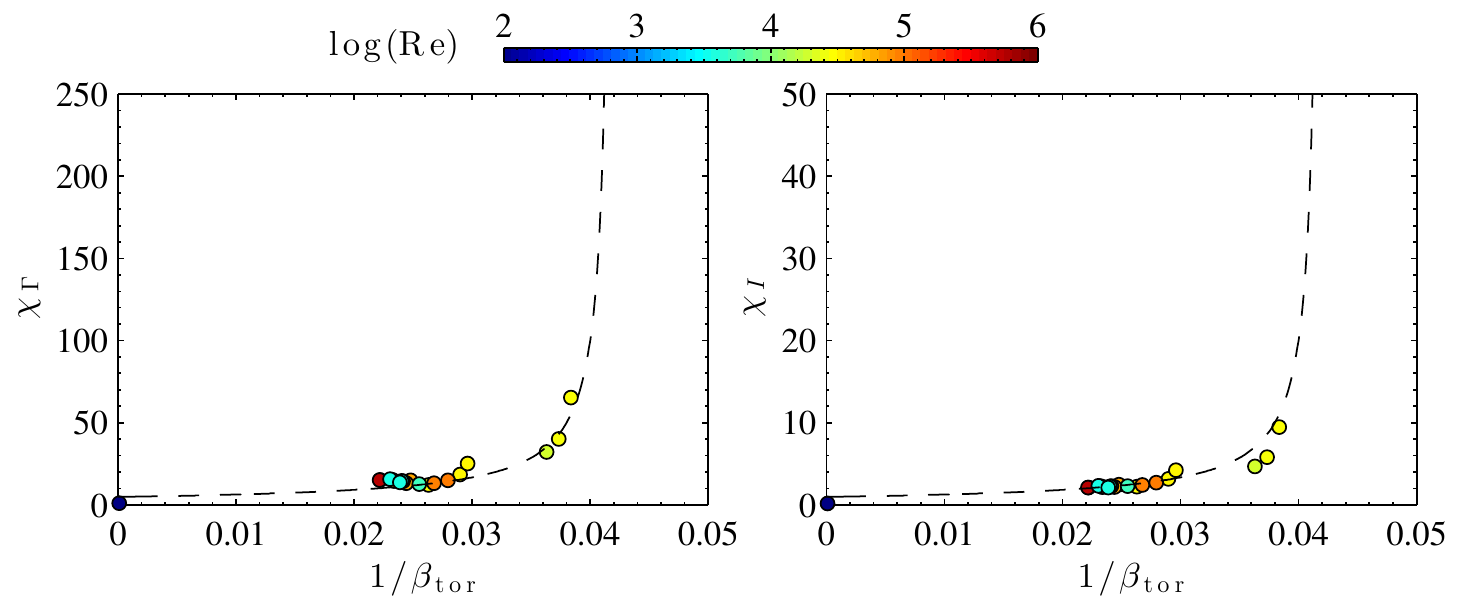}
\caption{The susceptibilities $\chi$ as a function of the temperature. The black dotted line is a fit $b/(1/\beta-1/\beta^\star)$, with $b=0.04$  for $\chi_I$ and $b=0.2$  for $\chi_\Gamma$ and $1/\beta_\star=0.044\pm0.03$.} 
\label{fig:Susceptibilities}
\end{figure}

\begin{figure}[!htb]
\centering
 \includegraphics[width=0.99\columnwidth]{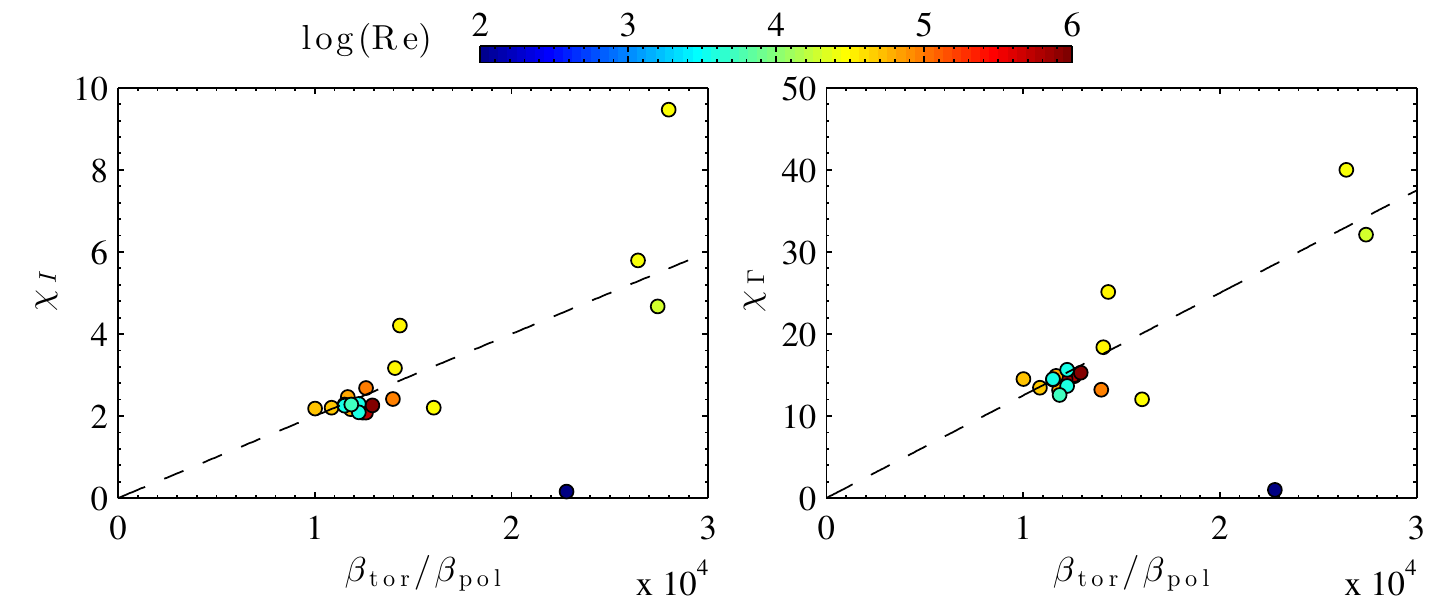}
\caption{Fluctuation-dissipation relation: the susceptibilities $\chi$ as a function of the $\beta_{\tor}/\beta_{\pol}$. Left: $\chi_I$. Right: $\chi_\Gamma$. The dotted line are linear fits, with respective slopes $10^{-3}$ et $2 \times 10^{-4}$. } 
\label{fig:SuscepFluct}
\end{figure}

\begin{table*}[ht]
%\begin{tabular}{cc}
%\parbox{12cm}
\centering
\begin{tabular}
{||c||c|c||}%
\hline \hline
   ``Ferro-Magnetic'' Quantity   & Hydrodynamic Analog & Name\\
\hline \hline
Spin & $(ru_\phi,\omega_\phi/r)$& ``Beltrami Spin''\\
%\hline
\multirow{2}{*}{Magnetization} &\multirow{2}{*}{${\bf M}=(I,\Gamma)$}   & \multirow{2}{*}{Angular momentum and circulation}\\
& & \\
%\hline
Thermostat & &Forcing and dissipation\\ %\hline
Temperature &$(1/\beta_{\tor},1/\beta_{\pol})$  &Fluctuations\\ %\hline
Symmetry breaking fields & $(\theta,\gamma)$ & Rotation and Torque numbers\\ %\hline
Susceptibility &$(\chi_I,\chi_\Gamma)$ &\\ %\hline
\hline \hline
\end{tabular}
\caption{Analogy between a two components spin system and the VK flow \label{Table:analog}}
\end{table*}
The phase transition is made apparent by the  study  of the behavior of the magnetization as a function of the temperature $1/\beta_\tor$. This is shown in Figure \ref{fig:Magnetization}, where the mean magnetizations at $\theta=0$ for all impellers (any $\alpha$) and Reynolds numbers have been gathered. The data collapse nicely on a well-defined curve. Compared to a ferro-magnetic system, the behavior is here reversed. A non-zero magnetization occurs when the temperature is higher than a critical temperature $1/\beta_{\tor}^{\star}({\bf M}) \simeq 0.044\pm0.03$. The data then suggest, that the magnetization grows as the square root of the distance to the critical temperature, \emph{viz.}, $|{\bf M}| \propto \sqrt{1/\beta_\tor - 1/\beta^{\star}_\tor}$, a behavior reminiscent of standard lattice models with mean-field interactions between the spins.
Similarly, the behavior of the susceptibility $\chi$ seems to exhibit a divergence around a critical temperature $1/\beta_\tor^\star(\chi) \simeq 0.044\pm0.03$ ---\, see Figure \ref{fig:Susceptibilities}. %
Because of the crucial role played by the poloidal fluctuations in the statistical theories described in Section \ref{sec:inviscid}, it is tempting to check whether some sort of fluctuation-dissipation relation holds,  involving both the poloidal and the toroidal fluctuations. Under its simplest expression, a fluctuation relation involving the susceptibility $\chi_\Gamma$ can be written as
\begin{equation}
\chi_\Gamma=\beta_\tor \int_\Dpiv \left(\avc{\omega_\phi^2}-\avc{\omega_\phi}^2\right) \propto \beta_{tor}/\beta_{pol}.
\label{fluctudissith}
\end{equation}
To check whether such a relation holds, we have plotted in Figure  \ref{fig:SuscepFluct}, $\chi_\Gamma$  against the quantity $\beta_\tor/\beta_\pol$. Indication of a linear trend is apparent. This is compatible with the observation made in the previous section, that the poloidal and the toroidal degrees of freedom are far more correlated in a VK set-up, than a purely axially symmetric interpretation of the dynamics would imply. This observation is in a sense compatible with the ``frozen-$\sigma$'' scenario, which tells that the poloidal degrees of freedom are somehow enslaved to the toroidal ones. Note that the linear trend also exists for $\chi_I$. In both cases, the prefactor is small (of the order of $1/1000$ to $1/20000$). In a Beltrami approximation, the ratio $\beta_\tor/\beta_\pol$ in inversely proportional to the mode numbers, \emph{i.e.} to the number of degrees of freedom \cite{naso2010statistical}. This remark would therefore provide an interpretation of the prefactor in the fluctuation-dissipation relation as the inverse number of degrees of freedom. This would give an estimate for the 
number of degrees of freedom of the order of $N\sim 10^3$ to $10^4$, much smaller than the traditional estimate $N=Re^{9/4}$ (that would rather give a number of the order of $10^{13}$ ).

\section{Conclusion}
In this work, we argued that the steady topologies observed in a turbulent VK set-up, as well as the transitions between those, could be interpreted  in terms of a statistical physics modeling. This allowed us to exhibit a deep analogy between the VK large scale dynamics and the standard behavior of Ferro-Magnetic material. 
We have shown that the topologies of the VK steady  states could be interpreted thermodynamically, using as a guideline an equilibrium theory that relies on the axially symmetric Euler equations and  an adequate modeling of the fluctuations. At first sight, there was no reason to expect that such an equilibrium theory 
could be relevant for the VK cases :  the lack of axial-symmetry for the instantaneous VK dynamics, the presence of a large-scale forcing which precludes any kind of separation of scale working hypothesis, the non-gaussianity of the VK fluctuations and the UV catastrophe associated to statistical theories based on the Euler axially symmetric equations are for example as many reasons which should imply that such an approach is bound to fail. \\
However, in the present case, the forcing and dissipation that are present in the experiment seem to play the role of  thermostats, which prescribe strong correlations between the degrees of freedom present in the flow. This allows for a statistical mean-field interpretation of the mean flow, provided one plugs into the theory additional ansatz non prescribed by the Euler equations.
In other words,  the VK steady states do \emph{not} correspond to ideal axially symmetric equilibria. Still, to first order, there exists some good indication that those can be interpreted as maximizers over an appropriate restricted subset of phase space of a suitably defined configuration entropy. 
This experimental fact provides an intuitive ground, to think about the VK experiment in terms of a statistical mean field theory.     
We have shown that this analogy goes beyond the mere description of the VK topologies. We showed that the transitions between the various topologies could be phrased in terms of phase transition and spontaneous symmetry breaking, shedding an unexpected analogy between VK turbulence and the theory of ferro-magnetism. \\
The empirical coincidence between the steady turbulent states and a class of ``meta-equilibria'' for the  Euler equation is puzzling and opens the question of how far "out-of-equilibrium" is our system. In traditional turbulence theory, the degree of  non- equilibriumness can be quantified by measuring the non-dimensional energy flux.  In the VK experiment, the latter is related to the torques exerted on both propellers.  One then observes that this quantity is minimal when the averaged topology is a two-cell  symmetric state, obtained by forcing at positive angle $\alpha$. These states appear to be better fitted by an equilibrium theory than the other bifurcated states, obtained at larger $\theta$ or for negatively curved propeller. It is also well known that the energy flux measured at the large scales of a turbulent flow is very close to zero. It may therefore well be that the large-scale topologies that we here considered  match a situation where the 
deviations from equilibrium  (quantified by the flux)  are sufficiently small, so that some kind of perturbative theory around the equilibria states may in the end be valid.

\ack
We thank the CNRS and the CEA for support, C. Herbert for interesting discussions  and P-P. Cortet and F.Ravelet for sharing the experimental data with us. 
This work received funding from  the European Community Framework Programme 7,(EuHit No.~Grant agreement no. 312778  and ERC Grant Agreement No.~240579), and from the French Agence
Nationale de la Recherche (Programme Blanc ANR-12-BS09-011-04).

\appendix
\section{Summary of axially symmetric statistical theories.}
\label{app:axiequilibria}

In this technical appendix, we explain how to obtain Table \ref{Table:equilibria}, in which the  outcomes of the different statistical theories based on the axially symmetric perfect fluid are summarized. Our point is not to discuss those theories very thoroughly, but to show how various ansatz for the poloidal fluctuations lead to different relations  between the average toroidal field and the average poloidal field.

Because of the specific shape of the inviscid invariants related to the ideal axially symmetric fluid, and in particular because of the long-range, mean field nature of the kinetic energy, the micro canonical averages can be computed  in terms of a local probability field : the macro state probability field.    The latter is defined though a local discretization of the physical domain combined with a  local averaging of the velocity field, the details of which can be found in numerous papers --- \, see for example \cite{potters2013sampling,thalabard2013statistical} for a recent exposition. 
In the case of the axially-symmetric perfect fluid,  it is convenient to define the macro state probability field  $p$ as 
\begin{equation}
 p_\br(\sigma,\xi) = \text{Proba}(ru_\phi=\sigma \text{\,and\,} \omega_\phi = r \xi \, \text{in the vicinity of  } \, \br). 
\end{equation}
The probability ``Proba'' that appears in the latter equation is a local micro canonical probability, which we later write $\langle \cdot \rangle$. 
In what follows, we use the short-hand notation $\int_{\sigma,\xi} = \iint_{\mbR^2} \d \sigma \d \xi$ and still denote $\int_\mD = (1/hR^2)\int_{r=0}^R  \int_{-h}^h r \d r \d z$.

To compute the axially symmetric equilibria, one first needs to translate the inviscid invariants  (\ref{eq:energyhelicity}) and  (\ref{eq:axicasimirs}) which by definition affect the microscopic dynamics,  into constraints for the macro-state probability fields. For axially symmetric flows, we write $\mA_{\sigma}$ (``toroidal areas'') , $\Gamma_{\sigma}$ (``partial circulations'') and $\mE$  the macro-state constraints induced by the axially-symmetric Toroidal Casimirs, Helical Casimirs and the energy respectively, \emph{viz.},
\begin{equation}
\begin{split}
\mA_{\sigma_0}[p] = \int_\mD \int_{\sigma,\xi} p_\br(\sigma,\xi) \delta(\sigma-\sigma_0),  \sieq & \Gamma_{\sigma_0}[p] = \int_\mD \int_{\sigma,\xi}\xi p_\br(\sigma,\xi) \delta(\sigma-\sigma_0), \tieq{and }\\
&\mE[p] =\int_\mD \int_{\sigma,\xi} \left\lbrace \dfrac{\psi \xi}{2} + \dfrac{\sigma^2}{2 r^2} \right \rbrace p_\br(\sigma,\xi),\\
\end{split}
\label{eq:macrostateconstraints}
\end{equation}
%
%
%The toroidal areas $\mA_{\sigma}$ and the partial circulations $\Gamma_{\sigma}$  are the macro state counterparts of the toroidal Casimirs and the generalized Helicities respectively. 
%
Another crucial macro state functional is  the ``macro state entropy'', defined by 
\begin{equation}
 \mS[p] = -\int_\mD \int_{\sigma,\xi}p_\br(\sigma,\xi)\log p_\br(\sigma,\xi).
 \label{eq:macrostateentropy}
\end{equation}
In principle, for a prescribed set of macro state constraints, say   $\alpha(\sigma)$, $\gamma(\sigma)$ and $E$,  the micro canonical averages $\langle \rangle$ can now be computed in terms of a most probable macro state probability field $p^\star$. This is a consequence  of an extension of Laplace's method of steepest descent called the Sanov theorem. $p^\star$  is then  obtained by maximizing the macro state entropy   (\ref{eq:macrostateentropy}) among the macro state probability fields that satisfy the prescribed set of macro-state constraints, namely : 
  \begin{equation}
p_\star = \arg \sup\left\lbrace \mS[p] |  \mA_\sigma[p]=\alpha(\sigma), \Gamma_\sigma[p]=\gamma(\sigma), \mE[p]=E \right \rbrace.
\label{eq:optimization}
  \end{equation}
In principle, the optimization problem  (\ref{eq:optimization}) can be solved explicitly in terms of Lagrange multipliers, and the solution glibly written as 
  \begin{equation}
\begin{split}
& p^\star_\br(\sigma,\xi) = \dfrac{1}{\mZ^\star_\br}\exp\left\lbrace
-\beta \dfrac{\xi \psi}{2} - \beta \dfrac{\sigma^2}{2r^2} +f(\sigma) + \xi g(\sigma) \right\rbrace, \tieq{with} \psi=-\dfrac{\mL^{-1} \langle\omega_\phi \rangle}{r} \\
& \tieq{and} \mZ^\star_\br = \int_{\sigma,\xi} \exp\left\lbrace-\beta \dfrac{\xi \psi}{2} - \beta \dfrac{\sigma^2}{2r^2} + f(\sigma) + \xi g(\sigma)\right\rbrace. %\tieq{and} \psi=-r^{-1}\mL^{-1} \langle\omega_\phi \rangle.
\label{eq:optimizationsolve}
\end{split}
  \end{equation}
The Lagrange multiplier $\beta$, and Lagrange functions $f$ and $g$ are determined by the (functional) derivatives of the partition function :   
\begin{equation}
 -\int_\mD \dfrac{\partial \log \mZ_\br}{ \partial \beta } = E, \sieq  \int_\mD\dfrac{\delta \log \mZ_\br}{ \delta f(\sigma_0)  } = \alpha(\sigma_0) \tieq{and }  \int_\mD\dfrac{\delta \log \mZ_\br}{ \delta g(\sigma_0)  } = \gamma(\sigma_0).
 \label{eq:Lagrange}
 \end{equation}
  The formulation  (\ref{eq:optimization})--(\ref{eq:Lagrange}) is however merely formal, as  the optimization problem (\ref{eq:optimization}) is in this case ill-defined. The trouble comes from  the poloidal degrees of freedom being in a sense not constrained enough by the macro state constraints of Equation (\ref{eq:macrostateconstraints}). The problem is apparent in the definition of the partition function (\ref{eq:optimizationsolve}) : the integral $\int_\xi$ there involved does  not in general converge. We think that this behavior is an avatar of the UV catastrophe encountered in the statistical theories of ideal 3D flows.
A phenomenological taming of the problem can be achieved by  further constraining the set of macro state fields over which the optimization problem  (\ref{eq:optimization}) is solved. This requires the use of additional \emph{ansatz}, some of which we below discuss.  
In order to carry out some explicit calculations and retrieve the equations of Table (\ref{Table:equilibria}), we will consider two simplified sets of macro-constraints (\ref{eq:macrostateconstraints}). 
\emph{(i)} In the \emph{Two-Level  modeling}, we prescribe the toroidal field to be a two-level, symmetric distribution, \emph{viz.}, 
$ \alpha(\sigma)  = p \delta(\sigma-1) +(1-p) \delta(\sigma+1)$. Only five constraints then remain from the set of constraints (\ref{eq:macrostateconstraints}) : the energy,  two toroidal areas $A_\pm$, and two partial circulations $\Gamma_\pm$. We write $f_\pm$ and $g_\pm$ the corresponding Lagrange multipliers that appear in Equation (\ref{eq:optimizationsolve}).
\emph{(ii)} In the \emph{Gaussian modeling}, we relax the constraint on the toroidal distribution, and only prescribe its average and variance computed over the domain. Similarly, instead of specifying the correlations between the poloidal and the toroidal field using the  partial circulations, we we only prescribe a value for the helicity and the circulation. We are then again left with five macro state constraints. As far as the optimization problem is concerned, this is tantamount  to prescribe  a quadratic $f$ and a linear function $g$  in Equation (\ref{eq:optimizationsolve_frozenr}), namely 
\begin{equation}
 f(\sigma)  = -\kappa \sigma^2 + \mu \sigma \tieq{and} g(\sigma) =  -h \sigma + \gamma.
\end{equation}
\subsection{Freezing the poloidal field : The ``frozen-\br'' scenario.}
A simple but extreme way to further constrain the optimization problem (\ref{eq:optimization}) is to assume  that the poloidal field at position $\br$ is prescribed and non fluctuating. We therefore solve (\ref{eq:optimization}) among the macro states of the kind: 
\begin{equation}
p_\br(\sigma,\xi) = p_\br(\sigma|\xi) \delta(\xi - \xi_0(\br)). 
\end{equation}
Whatever the field $\xi_0$, the macro state entropy  (\ref{eq:macrostateentropy})  now reads  $\mS[p] = -\int p_\br(\sigma|\xi_0)\log p_\br(\sigma|\xi_0)$. The extremal points of $\mS[p]$ are achieved for the now correctly defined macro states :  
\begin{equation}
\begin{split}
& p^\star_\br(\sigma,\xi) = \dfrac{1}{\mZ^\star_\br}\exp\left\lbrace - \beta \dfrac{\sigma^2}{2r^2} +f(\sigma) + \xi_0(\br) g(\sigma) \right\rbrace\delta(\xi - \xi_0(\br)),\\
& \tieq{where} \mZ^\star_\br = \int_{\sigma} \exp\left\lbrace - \beta \dfrac{\sigma^2}{2r^2} + f(\sigma) + \xi_0 g(\sigma)\right\rbrace\delta(\xi - \xi_0(\br)).
\label{eq:optimizationsolve_frozenr}
\end{split}
\end{equation}
An easy calculation then yields :
\begin{equation}
\begin{split}
\langle \sigma \rangle = \int_{\sigma,\xi} p_\br(\sigma,\xi) \sigma  & =(\mu-h \xi_0)/(2\kappa+2\beta/r^2) \tieq{in the Gaussian case,}\\
&= \tanh(A+B\xi_0(\br)) \tieq{in the two-level case.}
\end{split}
 \end{equation}
We write $D(\br)  = 2\kappa + 2\beta/r^2)$,  $A=(f_+-f_-)/2$ and $B=(g_+-g_-)/2$.  In both cases, we also obtain $\langle \xi \rangle  = \xi_0(\br)$, as indicated in Table \ref{Table:equilibria}.
\subsection{Enslaving the poloidal field :  the ``frozen-$\sigma$'' scenario}
Another way to remove the UV catastrophe is to prescribe that not only the poloidal averages conditioned on the toroidal areas but the whole conditional distributions $p(\xi|\sigma)$ are prescribed. We then choose to solve the optimization problem (\ref{eq:optimization}) over  macro states of the kind 
\begin{equation}
p_\br(\sigma,\xi) = p_\br(\sigma) p_\br(\xi|\sigma),   \tieq{with $p_\br(\xi|\sigma)$ being prescribed \emph{a priori}.} 
\label{eq:macrostate_frozen}
\end{equation}
The macro state entropy  (\ref{eq:macrostateentropy}) can now be written in terms of the (prescribed) entropy $s_\br(\sigma)$ of the conditional distributions as 
\begin{equation}
 \mS[p] = -\int_\mD \int_{\sigma}p_\br(\sigma)(\log p_\br(\sigma)-s_\br(\sigma)) \tieq{with} s_\br(\sigma)=-\int_\xi  p_\br(\xi|\sigma) \log p_\br(\xi|\sigma).
 \label{eq:macrostateentropy_frozen}
\end{equation}
Defining $\xi(\sigma) = \int_\sigma  \xi p_\br(\xi|\sigma)$, we find that the macro-states (\ref{eq:macrostate_frozen}) that solve the optimization problem and extremize $\mS[p]$ are those satisfying 
\begin{equation}
\begin{split}
&p_\br^\star(\sigma) = \dfrac{1}{\mZ_\br^\star}\exp\left\lbrace s_\br(\sigma) + \xi(\sigma) g(\sigma)+f(\sigma) - \beta \sigma^2/2r^2 - \beta \xi(\sigma) \psi/2\right\rbrace\\
&\text{with}\sieq  \mZ_\br^\star = \int_\sigma \exp\left \lbrace s_\br(\sigma) + \xi(\sigma) g(\sigma)+f(\sigma) - \beta \sigma^2/2r^2 - \beta \xi(\sigma) \psi/2\right \rbrace. 
\end{split}
\end{equation}
In the two-level case, we write $\xi_\pm = \int_\sigma  \xi p_\br(\xi|\sigma=\pm)$, and easily obtain  
\begin{equation}
\langle \sigma \rangle = \tanh (A_1+ B_1\psi) \tieq{and} \langle \xi \rangle = \dfrac{\sum_\pm \xi_\pm \exp \left \lbrace s_{\br,\pm} + \xi_\pm g_\pm +f_\pm - \beta \xi_\pm\psi/2\right \rbrace}{ \sum_\pm \exp \left \lbrace s_{\br,\pm} + \xi_\pm g_\pm +f_\pm - \beta \xi_\pm\psi/2\right \rbrace }.
\end{equation}
with $ A_1= (1/2)(s_{\br,+} + \xi_+ g_+ + f_+ -   s_{\br,-} + \xi_- g_- + f_-)$ and $B_1 = - \beta(\xi_+-\xi_-)/2$.

In the gaussian modeling, one needs to specify a shape for $\xi(\sigma)$ and the entropies $s_\br(\sigma)$ in order to carry the calculation further. If one assumes for simplicity $\xi(\sigma) = A(\br) \sigma +B$, and $s_\br(\sigma)=const.$, one readily obtains
\begin{equation}
\langle \sigma \rangle =  \dfrac{\gamma A  + \mu - h B - A\beta \psi/2}{2\left( \kappa + \beta/2r^2 +Ah \right)}\tieq{and} \langle \xi \rangle = A \langle \sigma \rangle +B.
\end{equation}
We therefore put $C=2(\gamma A  + \mu - h B)$, $D=4\left( \kappa + \beta/2r^2 +Ah \right)$ to obtain the result of Table \ref{Table:equilibria}.

\subsection{A regularization through  dissipation : the ``Gaussian'' scenario}
An alternative way to give some sense to the problem (\ref{eq:optimization}) is to restrict the maximization procedure to the macro state fields $p$'s whose local poloidal fluctuations are everywhere  bounded, namely the ones for which a prescribed $\lambda < \infty$ exists such that  $\int_{\sigma,\xi} \xi^2 p_\br(\sigma,\xi) < \lambda$.  One could rather wish to plug into the integral a $r^2$ coefficient so as to make this additional  macro-state constraint the exact counterpart to the poloidal viscous dissipative term made explicit  in Section (\ref{ssec:zeromode}), but this subtlety goes beyond the present point. 
In the gaussian modeling, the partition function logarithm of Equation (\ref{eq:optimizationsolve}) can be computed as  
\begin{equation}
 \log \mZ_\br = \log \pi - \dfrac{1}{2}\log \delta(\br) + \dfrac{\nu \mu^2 - h \mu (\gamma-\beta\psi/2) + (\kappa+\beta/2r^2) (\gamma-\beta\psi/2) ^2 }{4 \delta(\br)},
\end{equation}
where the Gaussian integration requires both $\nu$ and $\delta(\br) = (\kappa+\beta/2r^2)\nu-h^2/4$ to be positive.
We deduce 
\begin{equation}
 \langle \sigma \rangle = \dfrac{2\mu\nu - h \gamma +h\beta \psi/2}{4\delta(\br)} \tieq{and}  \langle \xi \rangle = \dfrac{2\gamma(\kappa+\beta/2r^2)  - h \mu -(\kappa+\beta/2r^2) \beta \psi}{4\delta(\br)}.
 \label{eq:gaussian_diss}
\end{equation}
Hence we retrieve the expression of Table \ref{Table:equilibria} by putting 
$A=4\mu\nu - 2h \gamma$, $B = h\beta$, $D = 8 \delta$, $C = 4\gamma(\kappa+\beta/2r^2)  - 2h \mu$ and  $F = -2(\kappa+\beta/2r^2) \beta$.
In the two-level case, the partition function reads 
\begin{equation}
 \log \mZ_\br = \dfrac{1}{2}\log \pi - \dfrac{1}{2}\log \nu + \log \sum_\pm \e^{f_\pm+(g_\pm-\beta\psi/2)^2/(4\nu) }.
\end{equation}
We deduce 
\begin{equation}
 \langle \sigma \rangle =  \tanh \left(A_1 + B_1 \psi\right)  \tieq{and}  \langle \xi \rangle = - \dfrac{\beta \psi}{4\nu} + \dfrac{1}{2 \nu} \dfrac{\sum_\pm g_\pm \e^{f_\pm - (g_\pm-\beta\psi/2)^2/2\nu} }{\sum_\pm e^{f_\pm - (g_\pm-\beta\psi/2)^2/ 2 \nu}},
 \label{eq:twolevel_diss}
\end{equation}
with $A_1 = (1/2) \left( f_+ -f_- +(g_-^2-g_+^2 )/(2\nu) \right)$, $B_1 = -\beta (g_- -g_+)/(4\nu)$.  To obtain the expression of Table \ref{Table:equilibria}, one needs to develop  the squares in the exponentials in both the numerator and the denominator of the expression for $\langle \xi \rangle$, to obtain $A_\pm = f_\pm - g_\pm^2/(2\nu)$ and $B_\pm = g_\pm \beta$.  

\subsection{An ideal microcanonical scenario}
Finally, in order to define a micro canonical measure for the axially symmetric Euler equations, the following strategy was  proposed in \cite{thalabard2013statistical} : \emph{(i)} regularize the optimization problem (\ref{eq:optimization}) by using a temporary cutoff, \emph{(ii)} solve the cutoff dependent optimization problem \emph{(iii)} Relax the cutoff dependency by letting the cutoff go to $\infty$ while using an appropriate scaling for the Lagrange multipliers. The result of \cite{thalabard2013statistical}  can be retrieved from the Gaussian scenario just described and equations (\ref{eq:gaussian_diss}) and (\ref{eq:twolevel_diss}). 
From Equation  (\ref{eq:gaussian_diss}), one prescribes $\beta \sim \beta_0 \nu$, $h\sim h_0 \nu$ , $ \gamma = \gamma_0 \nu$ as $\nu \to 0$  to obtain 
\begin{equation}
 \langle \sigma \rangle = \dfrac{\mu}{2 \kappa} \tieq{and}  \langle \xi \rangle = \dfrac{\gamma_0}{2} - \dfrac{h_0 \mu}{4\kappa} -\dfrac{\beta_0 \psi}{4}.
 \label{eq:gaussian_diss_asymp}
\end{equation}
From Equation  (\ref{eq:gaussian_diss}), one  prescribes $\beta \sim \beta_0 \nu$, $g_\pm \sim g_{\pm,0} \nu$, $ f_\pm \sim f_{\pm,0} $ as $\nu \to 0$, to obtain 
\begin{equation}
 \langle \sigma \rangle = \tanh\left( \dfrac{f_{+,0}-f_{-,0}}{2}\right)  \tieq{and}    \langle \xi \rangle = - \dfrac{\beta_0 \psi}{4} + \dfrac{1}{2} \dfrac{\sum_\pm g_{\pm,0} \e^{f_{\pm_0}} }{\sum_\pm e^{f_{\pm,0}}}.
 \label{eq:twolevel_diss_asymp}
\end{equation}
In both cases, the average toroidal field is constant and the average poloidal field is linear with $\psi$. Hence, we retrieve the expression of Table \ref{Table:equilibria} by putting $A = \mu/2\kappa$, $C=  \gamma_0/2 - h_0 \mu/4\kappa$, $B = B_1 = -\beta_0 \psi/4$, $A_1 =\tanh\left((f_{+,0}-f_{-,0})/2\right)$  and $C_1 = (1/2) \left(\sum_\pm g_{\pm,0} \e^{f_{\pm_0}} \right)/\left(\sum_\pm \e^{f_{\pm,0}} \right)$.

\section*{References}
\bibliographystyle{iopart-num}
\bibliography{ferro}
\end{document}